\documentclass[twocolumn, amsmath,amssymb,pra, showpacs, floatfix]{revtex4}
\usepackage[T1]{fontenc}
\usepackage{epsfig}
\usepackage{color}
\usepackage{graphicx}
\usepackage{bm}
\usepackage{epstopdf}
\usepackage{pdfpages}

\newcommand{\mket}[1]{\left|#1\right\rangle}

\newcommand{\mbra}[1]{\left\langle#1\right|}
\newcommand{\figscale}{0.5}

\begin{document}

\title{Computation of local exchange coefficients in strongly interacting one-dimensional few-body systems: 
local density approximation and exact results}

\author{O.~V. Marchukov}
\author{E.~H. Eriksen} 
\author{J.~M. Midtgaard} 
\author{A.~A.~S. Kalaee} 
\author{D.~V. Fedorov} 
\author{A.~S. Jensen} 
\author{N.~T. Zinner}
\affiliation{Department of Physics and Astronomy, Aarhus University, DK-8000 Aarhus C, Denmark}

\date{\today}

\begin{abstract}
One-dimensional multi-component Fermi or Bose systems with strong zero-range interactions can be described in terms 
of local exchange coefficients and mapping the problem into a spin model is thus possible.
For arbitrary external confining potentials the local exchanges are given by highly 
non-trivial geometric factors that depend solely on the geometry of the confinement 
through the single-particle eigenstates of the external potential. To obtain accurate
effective Hamiltonians to describe such systems one needs to be able to compute these
geometric factors with high precision which is difficult due to the computational 
complexity of the high-dimensional integrals involved. An approach using the local density approximation 
would therefore be a most welcome approximation due to its simplicity. Here we assess
the accuracy of the local density approximation by going beyond the simple harmonic oscillator
that has been the focus of previous studies and consider some double-wells of current experimental 
interest. We find that the local density approximation works quite well as long as the 
potentials resemble harmonic wells but break down for larger barriers. In order to 
explore the consequences of applying the local density approximation in a concrete 
setup we consider quantum state transfer in the effective spin models that one
obtains. Here we find that even minute deviations in the local exchange coefficients
between the exact and the local density approximation can induce large deviations
in the fidelity of state transfer for four, five, and six particles.
\end{abstract}

\maketitle

\section{Introduction}
The field studying atomic gases at extremely low temperatures has seen riveting developments over the
past decade \cite{bloch2008}. This has pushed the field into a rather unique position where cold atoms
in various geometries can be used to do quantum simulation of widely used models from other fields including
condensed-matter physics \cite{lewenstein2007,esslinger2010}, particle and high-energy physics \cite{son2007,strings2008}, nuclear 
physics \cite{zinner2013}, and even chemistry \cite{baranov2012}. A remarkably useful feature of cold atomic
gas experiments is the control over external trapping parameters that makes it possible to explore physics 
in different dimensionalities \cite{bloch2008}. In particular, a number of impressive experiments have 
studied various aspects of interacting bosons in one dimension \cite{moritz2003,stoferle2004,kinoshita2004,paredes2004,kinoshita2006,haller2009,haller2010}. 
Using the atomic interaction resonances caused by one-dimensional 
confinement \cite{olshanii1998} it has thus become possible to realize many interesting one-dimensional 
systems including the famous hard-core bosonic Tonks-Girardeau gas \cite{tonks1936,girardeau1960} (see
Refs.~\cite{kinoshita2004,paredes2004}). Most recently, one-dimensional systems of interacting fermions have
been also realized in experiments \cite{pagano2014}.

In the last few years it has become possible to control the particle number in one-dimensional 
experiments with great accuracy \cite{serwane2011}, allowing one to build a controllable few-body 
system of fermions and study fermionization for strong interactions \cite{zurn2012},
pairing \cite{zurn2013}, impurity physics \cite{wenz2013}, a two-site Hubbard model \cite{murmann2015a}, 
and Heisenberg spin models \cite{murmann2015b}. These developments have sparked a great deal 
of excitement in the community studying few-body physics and its relation to many-body phenomena.
Over the last decade a lot of new aspects of these one-dimensional systems have been covered by different authors. 
This includes the physics of small trapped bosonic systems (single- and two-component) 
\cite{zollner2005,deuret2007,tempfli2008,girardeau2011,brouzos2012,brouzos2014,wilson2014,zinner2014,garcia2015,dehk2015a,pietro2015}, 
details of few-fermion systems with two values of an internal degree of freedom (spin) 
\cite{guan2009,yang2009,rubeni2012,brouzos2013,bugnion2013,gharashi2013,sowinski2013,gharashi2014,artem2014,cui2014,loft2015,lundmark2015,gharashi2015,unal2015,artem2015f} and various aspects of the transition from few- to many-body 
physics \cite{girardeau2010,guan2010,astrak2013,volosniev2013,lindgren2014,deuret2014,volosniev2014,levinsen2014,sowinski2014,grining2015}. Furthermore, a number of studies have looked into mixed systems of bosons and 
fermions and systems with particles of unequal mass 
\cite{girardeau2004,girardeau2007,zollner2008,deuret2008,fang2011,harshman2012,garcia2013a,garcia2013b,harshman2014,campbell2014,garcia2014a,damico2014,mehta2014,garcia2014b,barf2015,artem2015b,dehk2015b,harshman2015,pecak2015,grass2015}.

In the present paper we are interested in studying one-dimensional 
fermions or bosons with strong repulsive short-range interactions.
As has been recently discussed, in the strongly interacting limit the 
system may be described by an effective Hamiltonian which has the 
form of an (anisotropic) Heisenberg spin model \cite{volosniev2013,deuret2014}.
It was predicted theoretically that by tuning the interaction strength 
from the weak (repulsive) to the strongly interacting regime for a 
two-component Fermi system in one dimension one may arrive in the 
ground state of an antiferromagnetic Heisenberg model \cite{volosniev2013}
and recently this was confirmed in experiments \cite{murmann2015b}. 
However, the Heisenberg model obtained is non-trivial in the sense that 
the exchange couplings are determined by the local geometry of the 
trapping potential. Computing the local exchange coupling constants is
a formidable numerical task and thus several papers have discussed the 
possibility to use various approximations to obtain these quantities. 
A very neat approximation is the strong-coupling ansatz of Levinsen {\it et al.}
\cite{levinsen2014,pietro2015}
which allows one to get an extremely accurate set of exchange coefficients
for arbitrary system sizes in the case where the external confinement 
is given by a harmonic oscillator. It has also been conjectured that for 
smooth potentials, the well-known local density approximation (LDA) should give
a very accurate value for the local exchange coefficients \cite{deuret2014,levinsen2014}
and numerical results for up to six particles show that the deviations are
at most a few percent \cite{deuret2014}. This is a very reasonable 
expectation and many studies in the past have used the LDA when 
studying the properties of many-particle systems in one dimension 
\cite{petrov2001,recati2003,astrak2004,tokatly2004,astrak2005,orso2007,liu2008,colo2008,rosi2015}.
However, this does not necessarily imply that the LDA works equally well 
for smaller systems as large system sizes can sometimes average out some
of the finer details that the LDA may not capture.

In the present paper we test the performance of the LDA for strongly 
interacting one-dimensional systems with up to six particles by 
comparing it to exact calculations. We do this for several different 
potential forms, including some experimentally relevant double-well
geometries. To the best
of our knowledge, no previous study has presented results for 
the effective spin models with up to six particles in non-harmonic 
confinement. We gauge the performance of the LDA against exact 
results not only for statics (producing the local exchange constants
needed for the effective Hamiltonian) but also for dynamics. For 
the latter case we study quantum state transfer using the spin models
that one obtains with the LDA and with exact calculations. Transfer of 
quantum states in two-component spin systems is a delicate process that 
depends sensitively on the local exchange couplings and thus provides a
difficult challenge for the LDA. The dynamical propagation 
of information and correlations in one-dimensional setups with cold 
atoms is a focal point of research at the 
moment \cite{cheneau2012,fukuhara2013,hild2014,fukuhara2015}
and it is thus 
theoretically important to have accurate models for these systems also 
in the strongly interacting regime 
where time-dependent exchange and non-equilibrium quantum magnetism can be
studied \cite{hild2014,trotzky2008,artem2015f}. 

The paper is organized as follows. In Sec.~\ref{formal} 
we outline the model and its assumptions, and then discuss how 
to compute exchange coefficients exactly and within the LDA. Section~\ref{results} presents a 
comparison of exact and LDA results for double-well potentials.
In Sec.~\ref{qst} we introduce the spin model picture of strongly 
interacting 1D systems and we apply the local exchange coefficients
in the context of quantum state transfer to investigate some 
consequences of the differences between different computational 
schemes. Section~\ref{summary} contains a summary, discussions, 
and outlook.

\section{Formalism}\label{formal}
We consider particles that are confined to move in one dimension (1D)
and assume that along the direction of motion there is a 
trapping potential, $V(x)$, which is the same for all the 
particles. The particles have short-range interactions
that we model by a Dirac delta-function and in turn the 
Hamiltonian may be written in the following way
\begin{equation}
\label{hamiltonian}
H = \sum_{i=1}^N \left [ \frac{p_i^2}{2m} + V(x_i) \right ] + g_{1D} \sum_{i > j} \delta(x_i - x_j),
\end{equation}
where $p_i$ and $x_i$ are the momentum and coordinate operators of particle $i$, $m$ is the
mass of a particle and $V(x_i)$ is the trapping potential for the $i$th particle. 
The interaction strength is parametrized by $g_{1D}$. In experimental setups the 
1D confinement is achieved by applying a very tight transversal trap. It may
then be shown that $g_{1D}$ can be directly related to the scattering properties
of the non-trapped atoms and is a function of the low-energy three-dimensional 
scattering length $a_{3D}$ and the transverse trapping length $a_\perp$ \cite{olshanii1998}.
What is very interesting is that one finds resonances where $g_{1D}$ diverges
due to the presence of the transverse confinement. This has been clearly demonstrated
in recent experiments \cite{zurn2012}. In particular, the regime where 
$|g_{1D}|$ is very large is accessible experimentally \cite{murmann2015b}. 
For simplicity we will use the notation $g=g_{1D}$ from now on since this 
can give rise to no ambiguities in the present context.

In the present paper we will consider two-component Fermi systems with 
$N = N_\uparrow + N_\downarrow$ atoms,
where $N_\uparrow$ and $N_\downarrow$ are number of atoms of components 
with spin projection up and down, respectively. The zero-range two-body
interaction in the Hamiltonian will act only between pairs with opposite
spin projection. For pairs with the same spin projection the Pauli principle
requires antisymmetry upon exchange of the two particles. As the wave function 
must also be continuous, we may infer that when two particles with identical 
spin projections are close to each other the wave function has a continuous 
first derivative. In turn, the Dirac delta-function two-body interaction 
has no effect. 
In experiment with atoms there are interactions between pairs of atoms with
identical spin projection, but they are highly suppressed due to their 
short-range nature and thus can be safely neglected for our purposes.
The general Hamiltonian above takes this implicitly into account as
we ensure antisymmetry among like components in our $N$-body 
wave functions. We note, however, that many of our results may
be easily transferred to two-component bosons with uniform 
interactions, i.e. a single $g$ controlling the interactions 
between both identical and different components \cite{volosniev2014}.

We now consider the strongly interacting regime where $1/g\to 0$ 
\cite{volosniev2013, volosniev2014, zinner2014}. 
The most general eigenstate of the Hamiltonian in the 
limit $1/g\to 0$ is \cite{volosniev2014}
\begin{equation}
\label{eigenstate}
\Psi = \sum_k a_k \theta(x_{P_k(1)}, \dotsc, x_{P_k(N)}) \Psi_0(x_1, \dotsc, x_N),
\end{equation}
where the sum is over the $N!$ permutations, $P_k$, of the coordinates, $a_k$ are the coefficients
that depend on the ordering of the particles, $\theta(x_1, \dotsc, x_N) = 1$, when $x_1 < x_2 < \dots < x_N$
and zero otherwise. $\Psi_0$ is a fully antisymmetric $N$-particle Slater determinant wave function constructed
from $N$ single-particle wave functions that are obtained by solving the 
corresponding single-particle Schr{\"o}dinger equation with the potential $V(x)$. In 
this paper we are interested in the lowest energy manifold of $N$-body states
which are obtained by taking the single-particle states with the $N$ lowest
energies. 

The wave function $\Psi_0$ describes
the non-interacting $N$-fermion system with the energy $E_0$. It is 
important to note that this $N$-body energy
is $M(N_\uparrow, N_\downarrow) = N! /(N_\uparrow!N_\downarrow!)$
times degenerate, and thus there is a manifold of $M(N_\uparrow,N_\downarrow)$
degenerate $N$-body states in the limit $1/g\to 0$ which is a quasi-degenerate
manifold at large but finite $g$ (which is where we will be working below as 
we consider quantum state transfer). This degeneracy arises from the fact that
in the $1/g\to 0$ limit the particles become essentially impenetrable. Yet, 
there are still $M(N_\uparrow,N_\downarrow)$ distinguishable ways that the particles may
be ordered on a line which all have the same energy. In practice one may
think of these various orderings of the spins along a line as a set of 
basis states \cite{volosniev2013,volosniev2014,deuret2014}.

In the limit $1/g \to 0$ we can write the $N$-particle energy, $E$, to the linear
order in $1/g$ as~\cite{zinner2014, volosniev2013, volosniev2014}
\begin{equation}
E = E_0 - \frac{1}{g} \frac{\sum_{j=1}^{N-1} A_j \alpha_j}{\sum_{k=1}^{M(N_\uparrow, N_\downarrow)} a^2_k},
\end{equation}
where $A_j = \sum_{k>j} (a_{j} - a_{k})^2$. The important observation is that 
there is an $a_k$-independent coefficient, $\alpha_j$, in this expression. It is 
a geometric factor that depends solely on the total number of particles $N$ and 
on the potential $V(x)$ and its single-particle
eigenstates. Remarkably, it does {\it not} depend on what system
one is considering as long as the interactions are strong, i.e. as long as we consider
the limit $1/g\to 0$. If one is able to compute $\alpha_j$ for a given $V(x)$, then 
this may be used to study multi-component Fermi or Bose systems with all possible 
combinations of internal components among the particles. The geometric factor
$\alpha_j$ can in a certain sense be thought of as the local exchange coupling 
in the system. This interpretation is very useful when mapping the system onto 
a spin model \cite{volosniev2013,volosniev2014,deuret2014,levinsen2014}. If 
one thinks of the $N$ particles sitting on a line, then the index $j$ on 
$\alpha_j$ corresponds to a pair of particles and the exchange coupling for 
that pair is proportional to $\alpha_j$. Since strongly interacting 1D 
systems are governed by exchange processes, the statics and dynamics of 
such systems is essentially dictated by the $\alpha_j$ coefficients and
we would thus like to compute these in as general circumstances as 
possible. Therefore, the coefficients $\alpha_j$ will therefore be the main focus
of our discussion. 

An exact expression for $\alpha_j$ was first derived in \cite{volosniev2013}
\begin{equation}
\alpha_j =  \frac{\tfrac{\hbar^4}{m^2}\int \prod_{i=1}^{N}\mathrm{d}x_i \theta(x_1, \dotsc, x_N) \delta(x_1 - x_j) 
\left(\partial \Psi_0\right)^{2}}{\langle\Psi\mid\Psi\rangle},
\label{eq:alphaExact}
\end{equation}
where the derivative in the integral denotes
\begin{equation}
\partial \Psi_0=\left[\frac{\partial \Psi_0}{\partial x_1}\right]_{x_1 = x_N},
\end{equation}
where one first takes the derivative of the $N$-body antisymmetric function
$\Psi_0$ with respect to $x_1$ and then subsequently sets $x_1=x_N$.
The exact expression is an $(N-1)$-dimensional integral and the numerical calculation of $\alpha_j$ is
by no means an easy task. It is therefore desirable to consider whether 
appropriate approximations can be made to access these quantities also
for larger values of $N$. An important observation was made in Ref.~\cite{deuret2014}
where it was noticed that for $N\leq 6$, a local density approximation 
can be used in computing $\alpha_j$ and this yields results that are
off by only a few percent for the benchmark case where $V(x)$ is a
simple harmonic oscillator potential. Using a highly accurate 
ansatz wave function for the $N$-body problem, it later became
possible to get a highly accurate approximation to $\alpha_j$
for the harmonic trap for any $N$ given as a ratio of quadratic polynomials in $N$ \cite{levinsen2014}.

Here we are concerned with the question of how well the local density
approximation (LDA) does for different potentials. We must therefore
define and discuss how the LDA may be applied in the context
of strongly interacting particles in 1D. This discussion closely 
parallels that of Ref.~\cite{deuret2014}. The main inspiration for the
LDA method in the present context comes from earlier work on the 
Hubbard model using the Bethe ansatz where Ogata and Shiba have shown 
that in the strongly interacting limit the spin and charge dynamics
decouple \cite{ogata1990}. The spin degrees of freedom may correspondingly
be described by a spin model of the Heisenberg type (we return to this 
later on) with an exchange coupling that is proportional to the 
third power of the density \cite{matveev2004,guan2007,matveev2008}.
This result is derived for a homogeneous system with periodic boundary
conditions which are the typical basic conditions needed to solve the 
Bethe ansatz equations \cite{ogata1990}. In order to transfer these 
results into the present context with non-homogeneous confinement 
of the 1D system, Ref.~\cite{deuret2014} suggested to use the 
density from the LDA in the expression for the exchange coupling 
in the spin model. This yields the expression
\begin{equation}
\label{eq:alphaLDA}
\alpha^{(LDA)}_i = \frac{\hbar^4 \pi^2}{3m^2} n_{TF}^3(X_i).
\end{equation}
Here $n_{TF}(X_i)$ is the 1D Thomas-Fermi density~\cite{deuret2014,levinsen2014}
which is given by
\begin{equation}
n_{TF}(x) = \frac{1}{\pi\hbar} \sqrt{2m(\mu - V(x))},
\label{TFdensity}
\end{equation}
where $\mu$ is the chemical potential of the system. Ref.~\cite{deuret2014}
proposed to 
calculate the Thomas-Fermi density 
in the center-of-mass positions $X_i$ of the $i$th and $(i+1)$th particles. 
The position $X_i$ can be
written as~\cite{deuret2014} $X_i = \frac{1}{2}\int \mathrm{d}x (\rho^{(i)}(x) + \rho^{(i+1)}(x))$,
where 
\begin{equation}
\rho^{(i)}(x) = \frac{\int \mathrm{d}x_1 \dots \mathrm{d}x_N \theta(x_1, \dotsc, x_N) \delta(x - x_i)\Psi_0}{\langle\Psi\mid\Psi\rangle}.
\label{eq:density}
\end{equation}
Taking this position makes sense in light of the interpretation of 
$\alpha$ as a local exchange coupling of a pair of particles and thus 
by symmetry one should take the density at their common center-of-mass which 
may be determined within the $N$-body system. One may argue that any 
discrepancy between the exact value of $\alpha$ and the LDA approximation 
can be alleviated by simply picking the right values of $X_i$. However, 
finding the right $X_i$ appears as difficult as calculating the exact
$\alpha$ directly. Note also that for large $N$ finding $\rho^{(i)}$
is an equivalently difficult task and thus one needs some method of 
determining densities accurately. Here we will ignore these problems 
and be concerned with how well the LDA performs in cases where we can compute
both the exact and the LDA expression for $\alpha$ easily. We therefore
work with $N\leq 6$ here.
For a commonly used harmonic oscillator
trapping potential $V(x) = \frac{1}{2}m \omega^2 x^2$ the Thomas-Fermi
density can be written down in a simple form
\begin{equation}
\label{LDAHarmOsc}
n_{TF}(x) = \frac{1}{\pi \hbar}\sqrt{2m(N\hbar\omega - \frac{1}{2}m \omega^2 x^2)}.
\end{equation}
In this case the approximation is rather accurate
compared to the exact calculation and the relative error is not larger than $7$ percent and
decreases rapidly with increasing particle number $N$~\cite{deuret2014}.
However, for potentials of more complicated shape the calculation 
of the center-of-mass positions becomes	much more difficult. 

In the following we discuss the coefficients $\alpha_i$ and
$\alpha^{(LDA)}_i$, with $i = 1, \dotsc, N-1$ for different potential 
profiles.
The numerical calculations of Eqs.~\eqref{eq:alphaExact} and~\eqref{eq:density} consist of two main parts. 
First, we 
construct the Slater determinant $\Psi_0$ with the eigenstates, $\psi_j(x_i)$, of a particle trapped in the potential $V(x_i)$.
We assume that our system is confined in a box potential with length $L$. To handle this numerically, we use the
normalized eigenstates 
\begin{equation}
\phi_n(x_i) = \sqrt{\frac{2}{L}} \sin{\frac{\pi n x_i}{L}}, 
\end{equation}
where $n$ is an integer,
as an expansion basis for the functions 
\begin{equation}
\psi_j(x_i) = \sum_n a^{(j)}_n \phi_n(x_i). 
\end{equation}
To evaluate integrals in Eqs.~\eqref{eq:alphaExact}
and~\eqref{eq:density} we utilize a recursive many-dimensional integration
procedure with the one-dimensional integrals being calculated using a standard trapezoidal integration
routine. Even though more advanced integration methods of course are available they are not readily available
for the case of nested integrals (integration limits that depend on subsequent integrations).
We find that the trapezoidal integration
converges rapidly with the relative error for the values of the coefficients not exceeding $10^{-5}$. We find that
$50$ basis vectors $\phi_n$ and $50$ integration points are sufficient to achieve this accuracy in a reasonable time.
In the following sections we dismiss the superscript $(LDA)$ for
simplicity and make sure that it is clear from the context of the 
discussion and in the figures whether we are discussing the exact
expressions for $\alpha$ or the LDA approximations.

\begin{figure}
\centering	
\begin{minipage}{\linewidth}
\centering
\includegraphics[scale=0.65]{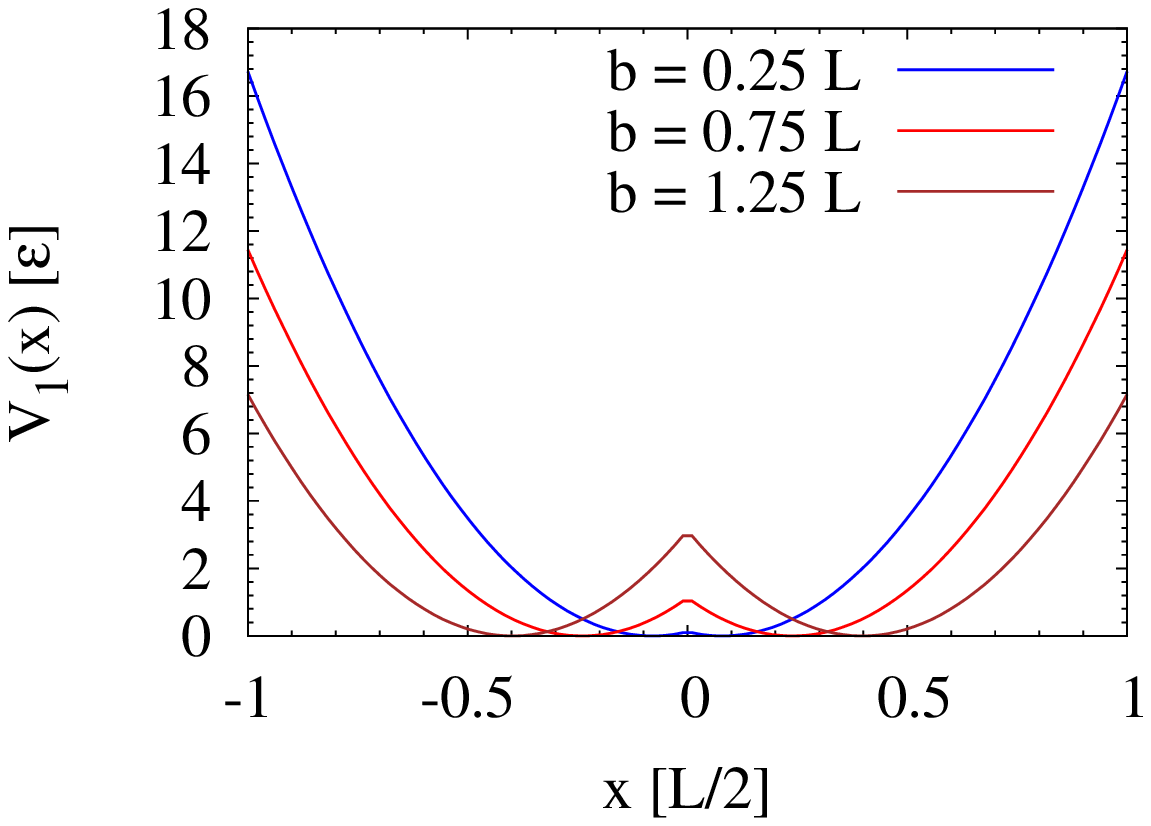}
\includegraphics[scale=0.65]{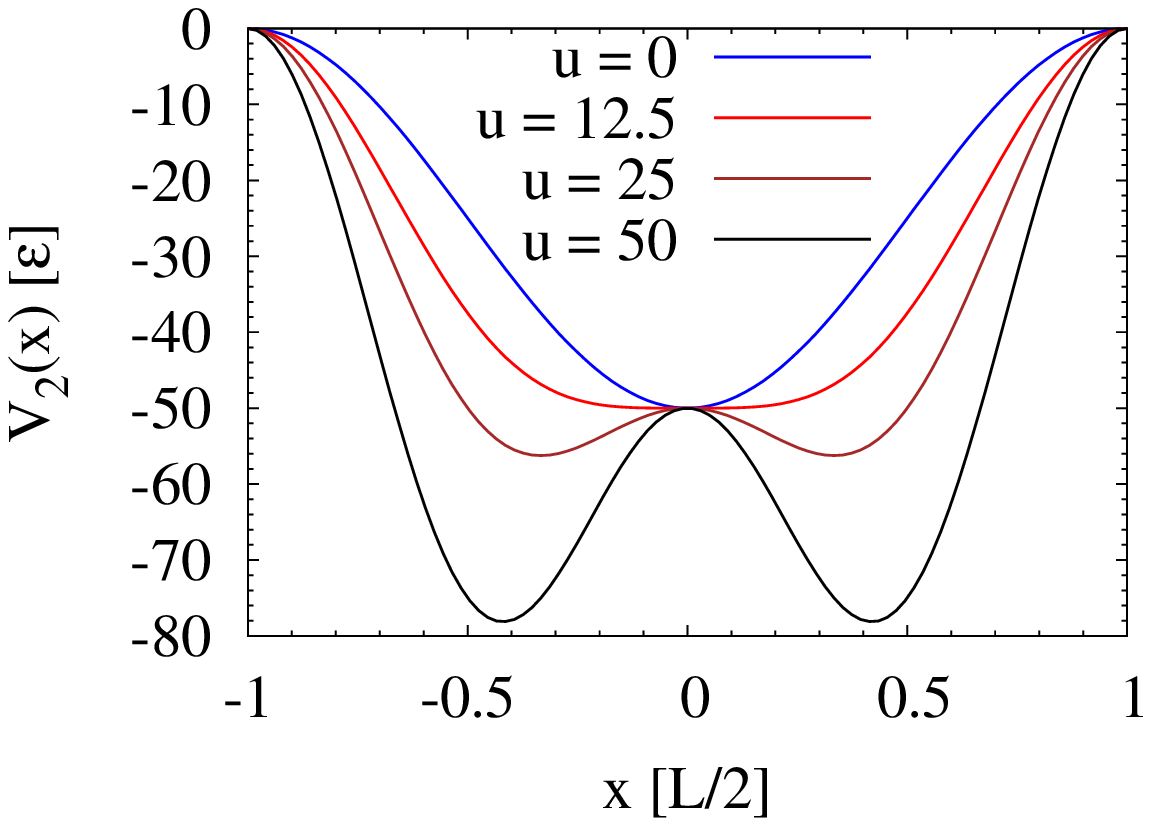}
\end{minipage}
\caption{The shapes of the potentials $V_1(x)$ from Eq.~\eqref{DW} (top panel) and $V_2(x)$ from Eq.~\eqref{AP} (bottom panel) for different values of parameters.}
\label{fig:potentials}
\end{figure}

\section{Comparison of exact and LDA coefficients}\label{results}
To perform a comparison
between the LDA results and the
exact solutions for the exchange coupling constants $\alpha_i$
we use two specific forms of a double well potential which have
flexibility to explore the similarities and differences between
the LDA and the exact calculation.
The first potential has the form
\begin{equation}
\label{DW}
V_{1}(x) = \frac{1}{2} k(|x| - b)^2,
\end{equation}
where $k$ and $b$ define the energy scale, 
and the height of the central barrier. Notice that this
potential has a cusp (discontinuity of the first derivative)
at $x=0$. This is perfectly well allowed in the formalism as 
this will not cause any troubling discontinuities in the 
single-particle wave functions or their first derivatives.
The second potential is a
symmetric trap of the form~\cite{volosniev2014}
\begin{equation}
\label{AP}
V_{2}(x) = - V_{0} \sin^2 \left [ \frac{\pi}{2} \left ( \frac{2x}{L} + 1 \right )\right ] - u\sin^2 \left [ \pi \left (\frac{2x}{L} + 1\right)\right ],
\end{equation}
where the values of $V_0$ and $u$ control the shape of the potential.
Figure~\ref{fig:potentials} shows the shapes of the potentials we use in the
article for different values of the control parameters.

\begin{figure}
\begin{minipage}{\linewidth}
\includegraphics[scale=\figscale]{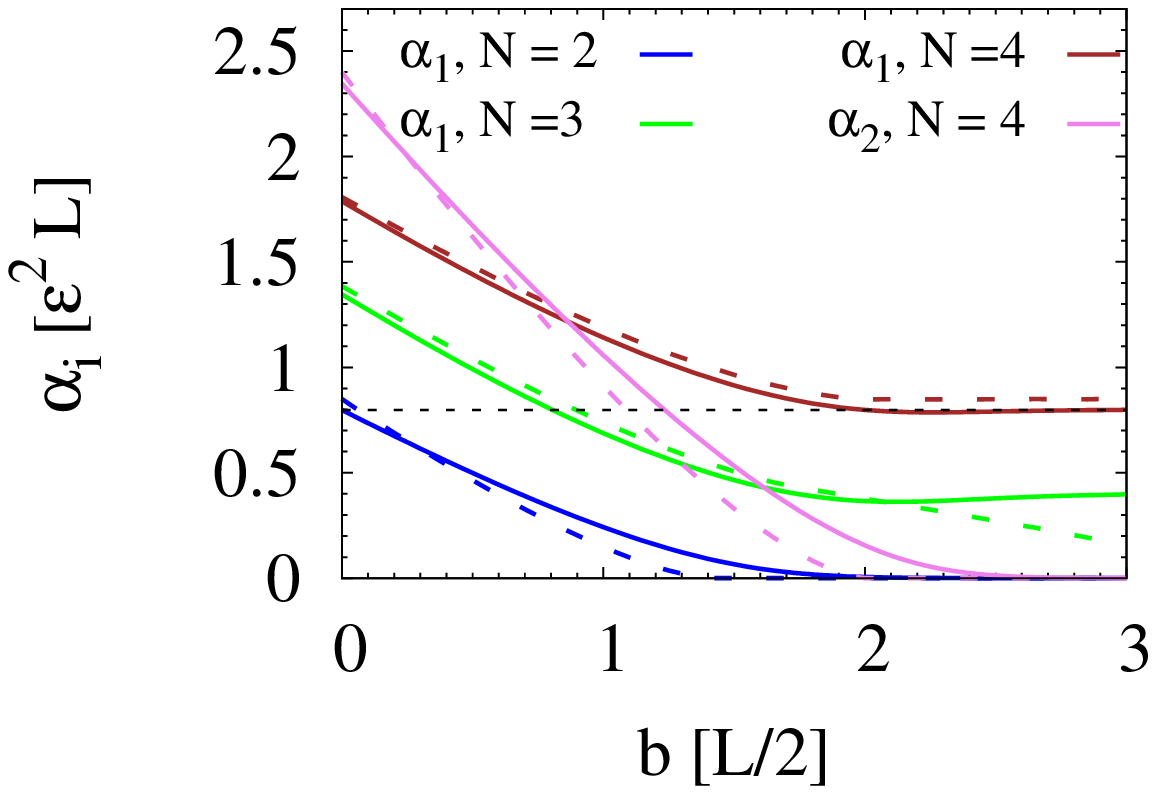}
\includegraphics[scale=\figscale]{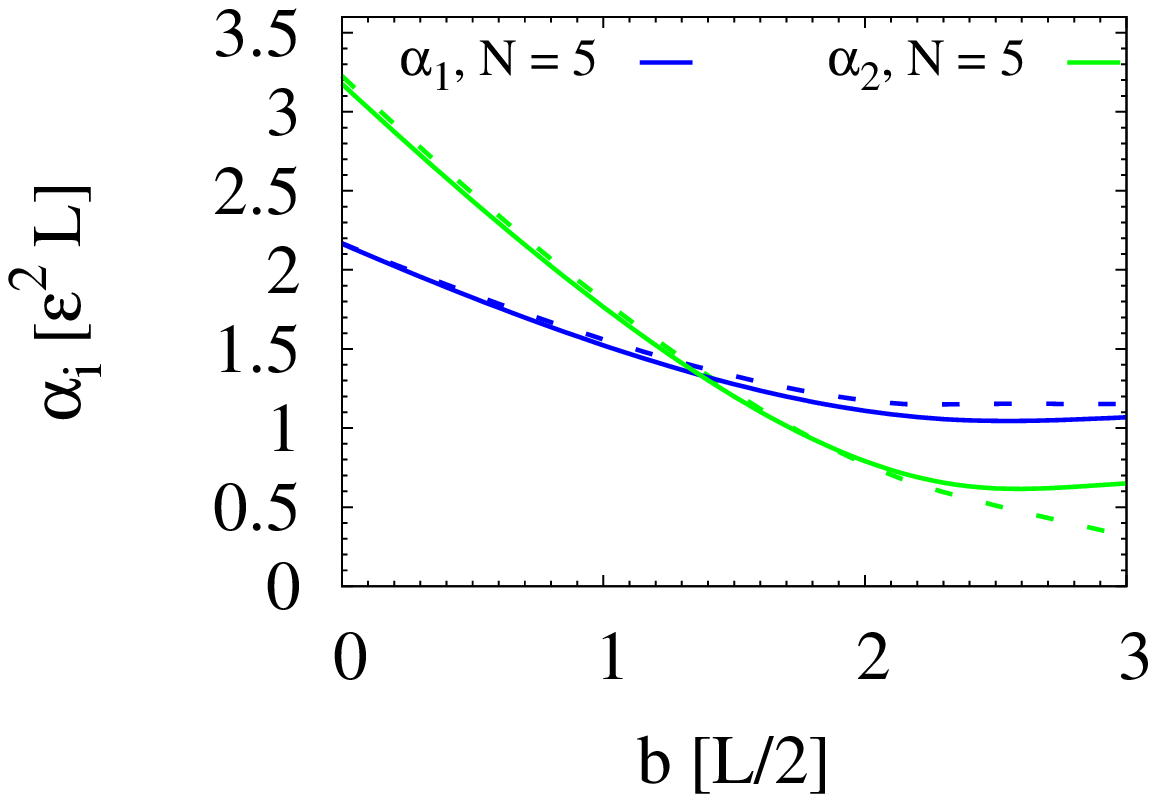}
\includegraphics[scale=\figscale]{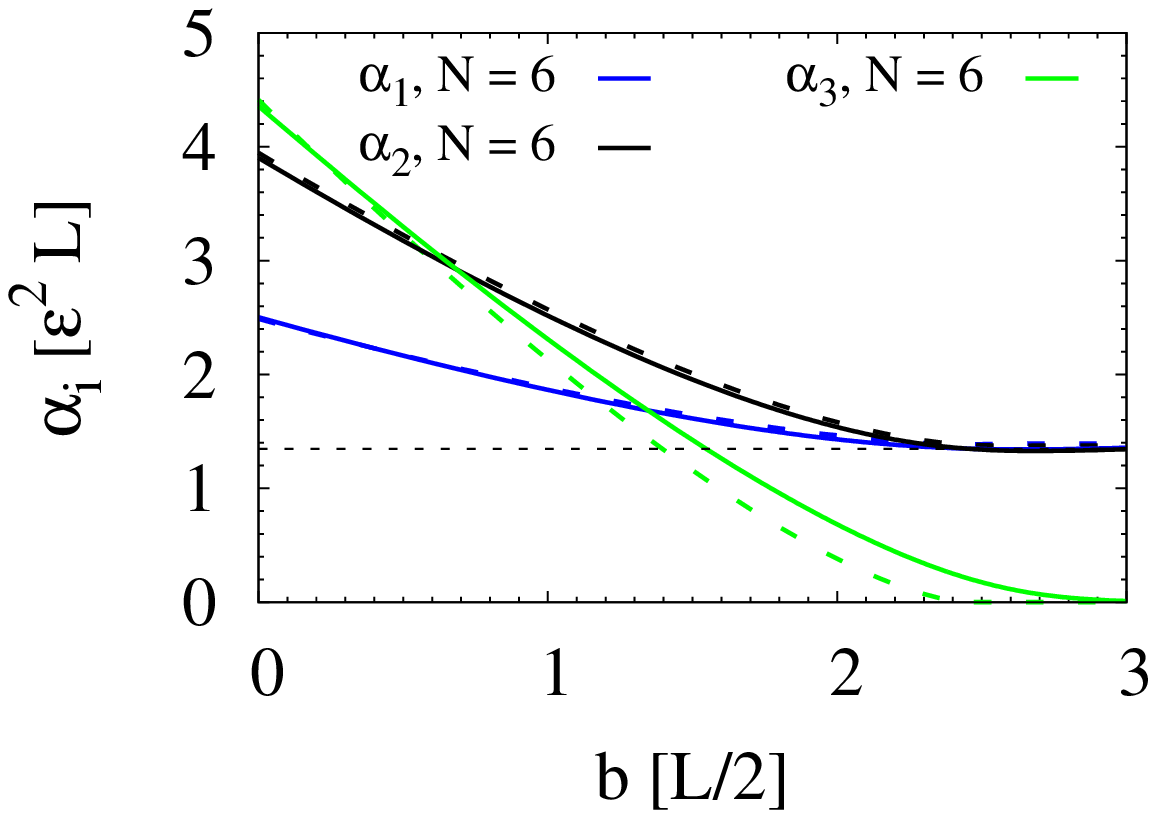}
\end{minipage}\hfil
\caption{Geometric exchange coupling coefficients $\alpha_i$ for the
double well potential in Eq.~\eqref{DW} with $k=1$ as function of the 
central barrier height $b$. The exact values based on the
expression in Eq.~\eqref{eq:alphaExact} are shown as solid lines, while the corresponding 
approximate values based on the LDA formula in Eq.~\eqref{eq:alphaLDA} are the dashed lines.
The top panels shows $N=2,3$ and 4, the middle panel has $N=5$, and the bottom panel has
$N=6$. Note that the parity symmetry of Eq.~\eqref{DW} implies that $\alpha_{N-i} = \alpha_i$. 
The dashed horizontal lines in the $N=4$ and $N=6$ 
cases correspond to the value $\alpha_1$ for $N=2$ and $N=3$, 
respectively. This is the excepted asymptotic values 
for large $b$ as discussed in the text.}
\label{fig:DWcomp}
\end{figure}

We note that we consider values in the interval $x \in [-L/2; L/2]$
where $L$ is the length of a box which confines the whole system. 
This is required for our numerical procedure where we use 
$L = 4 \pi$ in our calculations. 
Energies are measured in units of $\epsilon=4\hbar^2/m L^2$, 
with a factor of 4 coming from the box extending to $\pm L/2$
For the potential in Eq.~\eqref{DW} we set $k=1$ (in units of 
$L$ and $\epsilon$) while for Eq.~\eqref{AP} we have used 
$V_0=50$ in units of $\epsilon$. In deep enough potentials
the single-particle states do not feel the effect of this 
'outside' boundary. However, for shallow potentials some of the states
might reside in the whole box and thus be influenced by the outer wall.
We will see how it affects the exchange constants in the subsections below.

\subsection{Results}
\begin{figure}
\begin{minipage}{\linewidth}
\includegraphics[scale=\figscale]{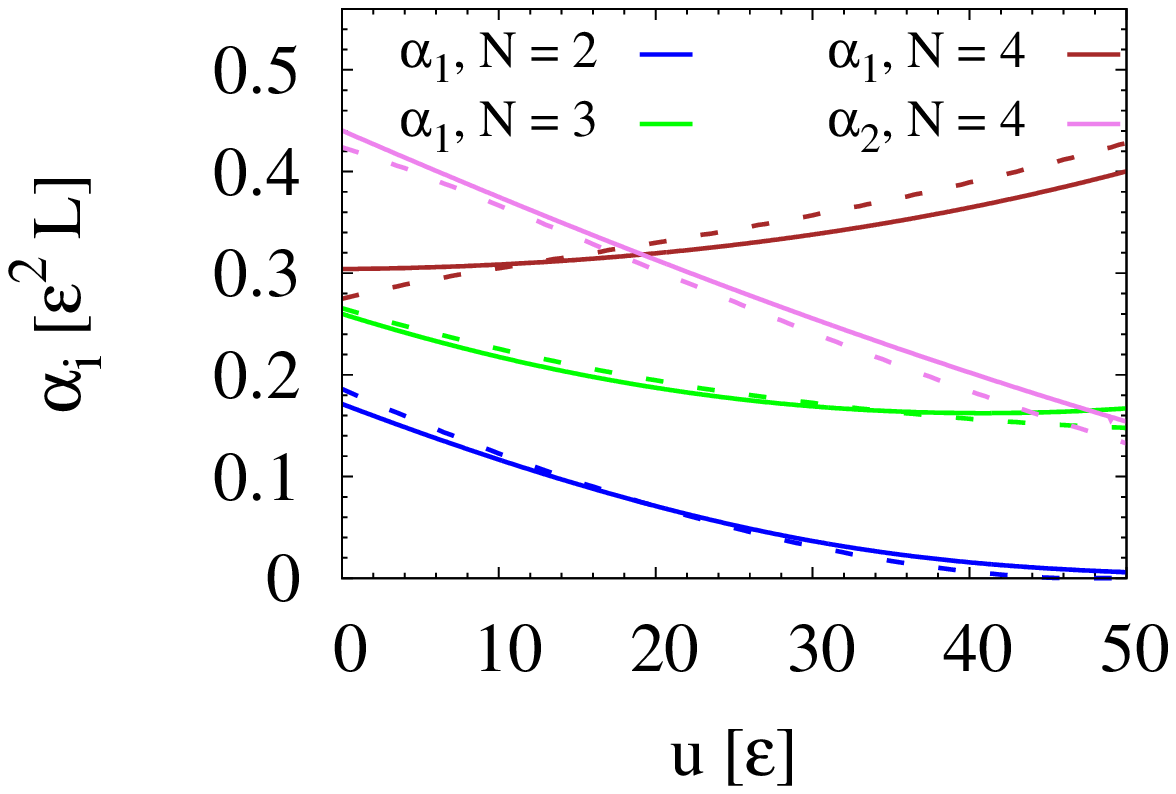}
\includegraphics[scale=\figscale]{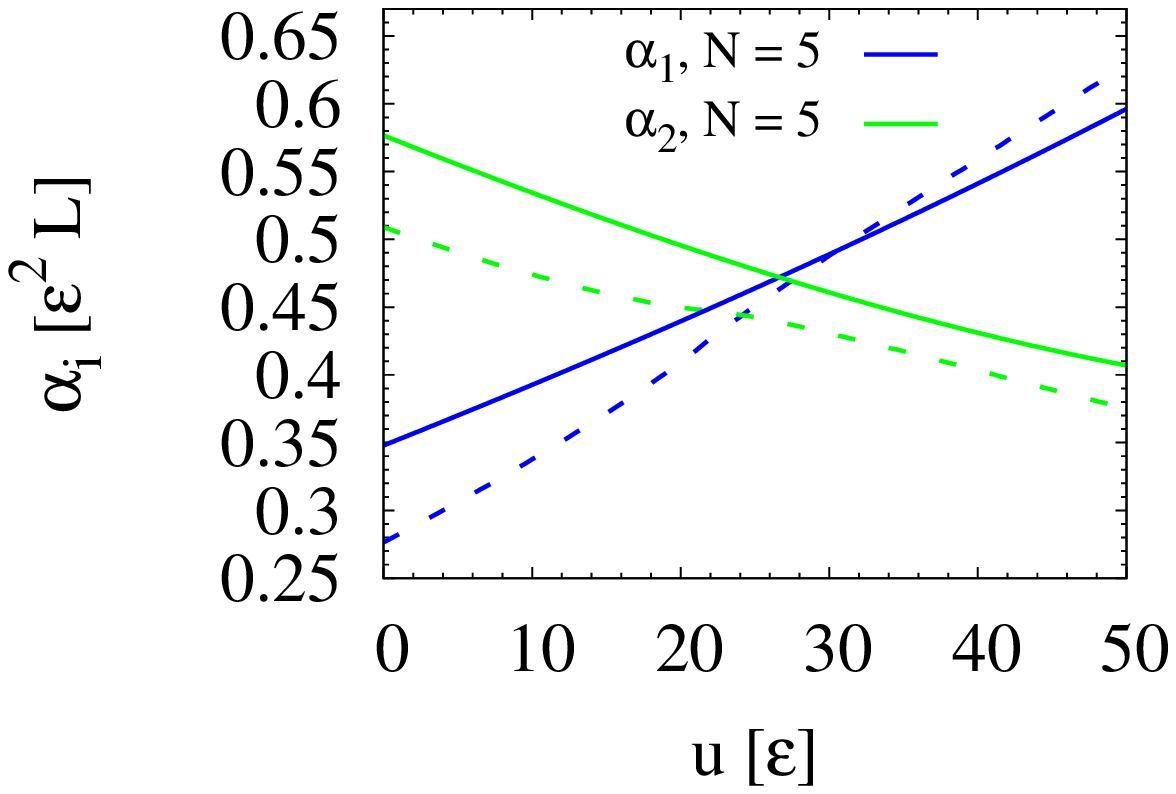}
\includegraphics[scale=\figscale]{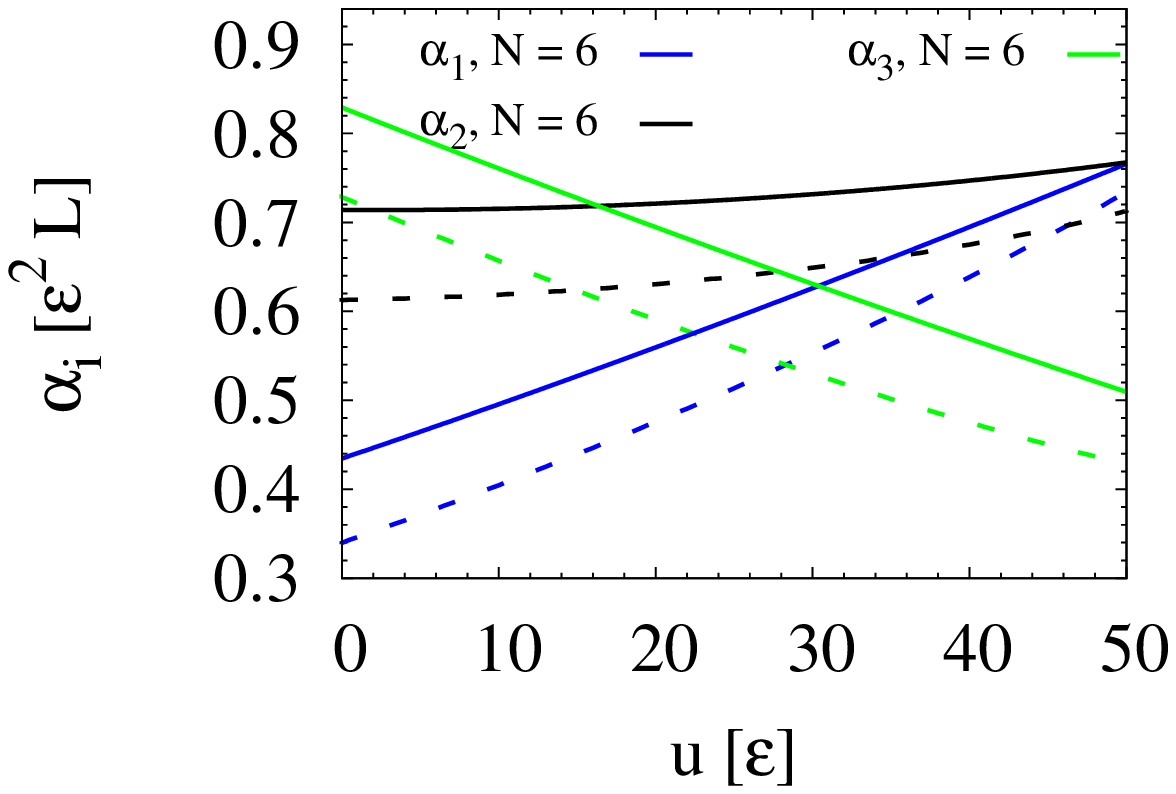}
\end{minipage}\hfil
\caption{
Geometric exchange coupling coefficients $\alpha_i$ for the
smooth double well potential in Eq.~\eqref{AP} as function of the parameter $u$
that controls the central barrier. Here we have set $V_0=50$. The exact values based on the
expression in Eq.~\eqref{eq:alphaExact} are shown as solid lines, while the corresponding 
approximate values based on the LDA formula in Eq.~\eqref{eq:alphaLDA} are shown as dashed lines.
The top panel shows $N=2,3$ and 4, the middle panel has $N=5$, and the bottom panel has
$N=6$. Note that the parity symmetry of Eq.~\eqref{AP} implies that $\alpha_{N-i} = \alpha_i$.}
\label{fig:APcomp}
\end{figure}

We now consider the coefficients $\alpha_i$ using Eqs.~\eqref{eq:alphaExact} and
~\eqref{eq:alphaLDA} for different values of the control parameters; the height of
the central barrier $b$ for the potential in Eq.~\eqref{DW} and the parameter $u$ for
the potential in Eq.~\eqref{AP}.
Figs.~\ref{fig:DWcomp} and~\ref{fig:APcomp} show both the LDA and exact
values of the coefficients $\alpha_i$ for different numbers of particles $N = 2-6$.
In this article we consider only spatially symmetric 1D potentials and so $\alpha_{N-i} = \alpha_i$.

In Fig.~\ref{fig:DWcomp} we show the results for the potential in Eq.~\eqref{DW}
comparing the exact results obtained from Eq.~\eqref{eq:alphaExact} to the LDA
results obtained with the formula in Eq.~\eqref{eq:alphaLDA} for particle numbers 
$N=2-6$. Note that we do not need to plot all the $\alpha_i$ coefficients as the 
parity invariance of the double well in Eq.~\eqref{DW} gives us the convenient 
symmetry relation $\alpha_{N-i}=\alpha_i$. The results demonstrate that the LDA
does very well for small values of the central barrier height $b$ which should not
be too surprising as the potential for small $b$ resembles very much that of a 
harmonic oscillator which was previously shown to be quite accurately described
within LDA \cite{deuret2014}. 
 
One notices that for the even particle numbers the middle coefficient, i.e. $\alpha_1$, 
$\alpha_2$, and $\alpha_3$ for $N=2$, $N=4$, and $N=6$, respectively, has similar behavior, 
and that this behavior is not captured well by the LDA for $b\gtrsim 0.75$.
In fact, this middle coefficient, $\alpha_{N/2}$,
calculated via Eq.~\eqref{eq:alphaLDA} goes to zero much faster than the exact value. 
This $\alpha_{N/2}$ pertains to the exchange coupling in the middle of the trap, i.e. 
to the exchange taking place right at the central barrier in the potential. It is therefore
determined by the probability for particle tunneling through this central barrier, and
we clearly see the LDA fail to capture this effect for larger barriers, i.e. larger values 
of $b$, where LDA underestimates exchange. This is connected to the fact that
LDA and in particular
the Thomas-Fermi density in Eq.~\eqref{TFdensity} is only well-defined in 
between the classical turning points, where $\mu - V(x) > 0$. Hence for large barriers 
where $\mu-V(x)\leq 0$, the Thomas-Fermi density goes to zero and thus in turn
the coefficient $\alpha_{N/2}$ goes to zero. For example, in the
case of the potential in Eq.~\eqref{DW} $\alpha^{(lda)}_{N/2} \equiv 0$ for $b > \sqrt{N}$ (when 
$k=1$ as we have here). The termination values of $b=\sqrt{2}$, $b=2$, and $b=\sqrt{6}$ can be clearly
seen on the horizontal axis for $\alpha_1$ ($N=2$) and $\alpha_{2}$ ($N=4$) in the 
top panel of Fig.~\ref{fig:DWcomp}, and for $\alpha_3$ ($N=6$) in the bottom panel of Fig.~\ref{fig:DWcomp}.

When the height of the barrier is large (large $b$) each of the two wells can be approximated by the harmonic oscillator
potential. We see that in this case the exact values of the coefficients becomes almost constant as function of the barrier height.
For even number of particles we intuitively expect that half of the particles will be located in each well. In the case with $N=4$ 
this would imply that $\alpha_1$ should go to a constant for large $b$, reaching the value corresponding to 
the $b=0$ case with $N=2$ (two particles in a single harmonic well). This is clearly seen in the top panel 
of Fig.~\ref{fig:DWcomp} where the horizontal dashed line marks the latter value. Likewise, for $N=6$, we expect a split into two three-body systems and thus that 
both $\alpha_1$ and $\alpha_3$ should be approaching the $\alpha_1$ value for $b=0$ and $N=3$. Again this 
is seen very nicely in the bottom panel of Fig.~\ref{fig:DWcomp} the first two coefficients for 
$N=6$ approach the dashed horizontal line corresponding to $N=3$ and $b=0$ (notice that there 
are different vertical scales on the three panels in Fig.~\ref{fig:DWcomp}). 

For the spatially symmetric potentials considered here with an odd number of particles, we cannot use the same logic 
of division of the particles into the two wells as the barrier grows large as for even particle numbers.
The overlap across the barrier of the single-particle wave functions will not vanish for odd particle numbers. 
Hence, the coefficients $\alpha_i$ which depend solely on these single-particle wave functions will also not 
vanish. We clearly see in the top and the middle panel of Fig.~\ref{fig:DWcomp} that for $b\gtrsim 2$ 
the LDA result does not do a good job in describing the exchange couplings. Again, this is caused by
the fast decrease of the Thomas-Fermi density on which the LDA result relies as the barrier increases. 
This decrease of density clearly takes place much faster than seen in the exact results. We thus see
that odd-even effects can be considerable in the comparison of the LDA and the exact method.

In order to further explore the case with odd particle numbers we have tried to apply a small tilt
to the potential in order to explicitly break the parity invariance. This could for instance
be done by an additional term that is linear in $x$ in the potential~\eqref{DW}.
When this is done for an odd number of particles the values of the exchange couplings approach
the values of the couplings of a smaller system. For instance, for the system consisting of $N=5$
particles, two of the particles will be located in one of the wells, with the interacting coefficient approaching the value
of $\alpha_1$ for two particles in a harmonic trap, and the other three will occupy 
the other well, with $\alpha_1 = \alpha_2$ approaching
those of a three-particle system in a harmonic trap. 
At the same time we find the intuitive result, namely that 
the interaction between these two subsystems will approach
zero as the barrier increases. We will not discuss the introduction 
of slight symmetry-breaking terms any further here. 
All in all, for the double well potential~\eqref{DW} the LDA approach provides a reliable approximation
to the exact values of the coefficients $\alpha_i$ for small central 
barrier heights (small values of $b$) and then becomes poor for 
larger barrier heights. An exception is found for even particles
numbers where the system splits into two equal size groups that 
can be described as particles in two separate harmonic wells. 
Here LDA does approach the exact results for the split system
asymptotically.

To further explore the robustness and/or failures of applying the LDA to 
our problem system, we now change the potential into a different kind of 
double well shape which has the form given in Eq.~\eqref{AP}. What one should 
notice here is that the potential in Eq.~\eqref{AP} is in a 
sense shallow, i.e. it does not increase to infinity around its edges. 
Therefore the box potential that we have as a hard-wall boundary around the system 
could be felt by the particles. 
This is in fact the case for larger particle numbers as we will now discuss. 

\begin{table}
\begin{center}
\begin{tabular}{| l | l | l |}
 \hline
 $N$&$\alpha_i^{(EXACT)} [\epsilon^2 L]$ & $\alpha_i^{(LDA)} [\epsilon^2 L]$\\ \hline \hline
 $2$ & $0.02487$ & $0.01328$ (47\%) \\ \hline
 $3$ & $0.06963$ & $0.04485$ (36\%) \\ \hline
 $4$ & $0.14921$ & $0.10630$ (29\%) \\ \hline
 $5$ & $0.27355$ & $0.20798$ (24\%) \\ \hline
 $6$ & $0.45260$ & $0.35911$ (21\%) \\
 \hline
\end{tabular}
\caption{Comparison of the exact calculation, $\alpha_{i}^{(EXACT)}$ (first column), to 
that of the LDA, $\alpha_{i}^{(LDA)}$ (second column), for a potential consisting of a 
flat box with infinite walls. Notice that in this case the exchange coefficients 
are independent of $i$, i.e. $\alpha_i=\alpha$. The percentages in parenthesis in the 
second column give the deviations of the LDA from the exact results. Notice that 
the LDA always underestimates the exchange coupling coefficients. \label{tab:box}}
\end{center}
\end{table}

First we consider the cases with $N=2$ and $N=3$ as shown in the top panel 
of Fig.~\ref{fig:APcomp}. Here we see that the qualitative behavior of the 
$\alpha_i$ coefficients is similar to the potential of Eq.~\eqref{DW} 
(shown in the top panel of Fig.~\ref{fig:DWcomp}) with the two-particle 
case shown a steady decrease with the barrier parameter, $u$, while the 
three-particle case first decreases and then has a slight increase again 
at large values of $u$. This is analogous to the behavior in the top panel 
of Fig.~\ref{fig:DWcomp}. We also notice that the LDA does a very good job
for the two- and three-particle systems for most values of $u$ in the 
top panel of Fig.~\ref{fig:APcomp}. For the $N=3$ case this should be 
compared and contrasted to the case in the top panel of Fig.~\ref{fig:DWcomp}
where the same $N=3$ case for the potential in Eq.~\eqref{DW} demonstrates 
that the LDA deviates significantly from the exact result 
for larger barrier heights (large $b$ values in that case). 

For more particles ($N\geq 4$) the LDA results start to deviate significantly 
from the exact results as we see in all three panels in Fig.~\ref{fig:APcomp}. 
We see that already for four particles the coefficient $\alpha_i$
differs significantly in the LDA and exact calculations. This difference only increases 
for larger particle numbers. The reason for this discrepancy can be traced back 
to the form of the potential in Eq.~\eqref{AP} and its shallow nature as compared 
to that in Eq.~\eqref{DW}. For the parameters we have chosen, the potential does
not become deep enough for five or six single-particle eigenstates to be confined
completely inside of the potential profile. This can be rephrased by stating that
not all the five or six lowest single-particle energy eigenvalues are in fact
negative. The most energetic single-particle wave functions that we require to 
build the totally antisymmetric function, $\Psi_0$, are thus feeling the outside
of the potential. In particular, they are influenced by the box potential that 
surrounds our system. While one may alleviate this problem by going to deeper
potentials (increasing $V_0$ in Eq.~\eqref{AP}) we choose $V_0=50$ in order
for us to study quantum state transfer in the last part of our paper and make 
contact with recent results that utilize the 
same parameters \cite{volosniev2014}.

To gain further insights into the influence of the box and the behavior of 
LDA in this respect we may consider just the box potential on its own.
In Tab.~\ref{tab:box} we present the results of applying both the exact
formula (first column) and the LDA version (second column )to the box potential.
We clearly see a large discrepancy for the particle numbers we have studied. 
The deviation of the LDA from the exact result is given in percentages in 
the parenthesis in the second column of Tab.~\ref{tab:box}. By doing a 
crude fit to the percentages we find that the deviations scale 
approximately with the particle number as a power law $N^{-0.73}$. We thus
find a quite slow convergence and for smaller system sizes the deviations
can be significant. 
We can see that as 
the single-particle wave functions are distributed over the whole confining infinite square well potential
the local density approach becomes rather inadequate. The 
large discrepancies in the values of the LDA and the exact results for 
the $\alpha_i$ coefficients of the shallow potential in Eq.~\eqref{AP} for 
particle numbers $N\geq 5$ can now 
be better understood. They are in large part the result of the fact that the 
outer box boundaries do become important for these particles numbers and 
that the LDA does a poor job of describing particles in a box with a flat
bottom.
We do see
that the LDA approximation gives qualitatively similar results to the exact calculation even for $N > 4$, but
quantitatively the LDA may differ substantially. We have checked that as 
one increases the depth of the potential in Eq.~\eqref{AP} one does indeed
see better agreement. However, the deviations we identified as common 
for both potential profiles within the LDA remain.

\begin{figure}
\begin{minipage}{0.9\linewidth}
\includegraphics[scale=\figscale]{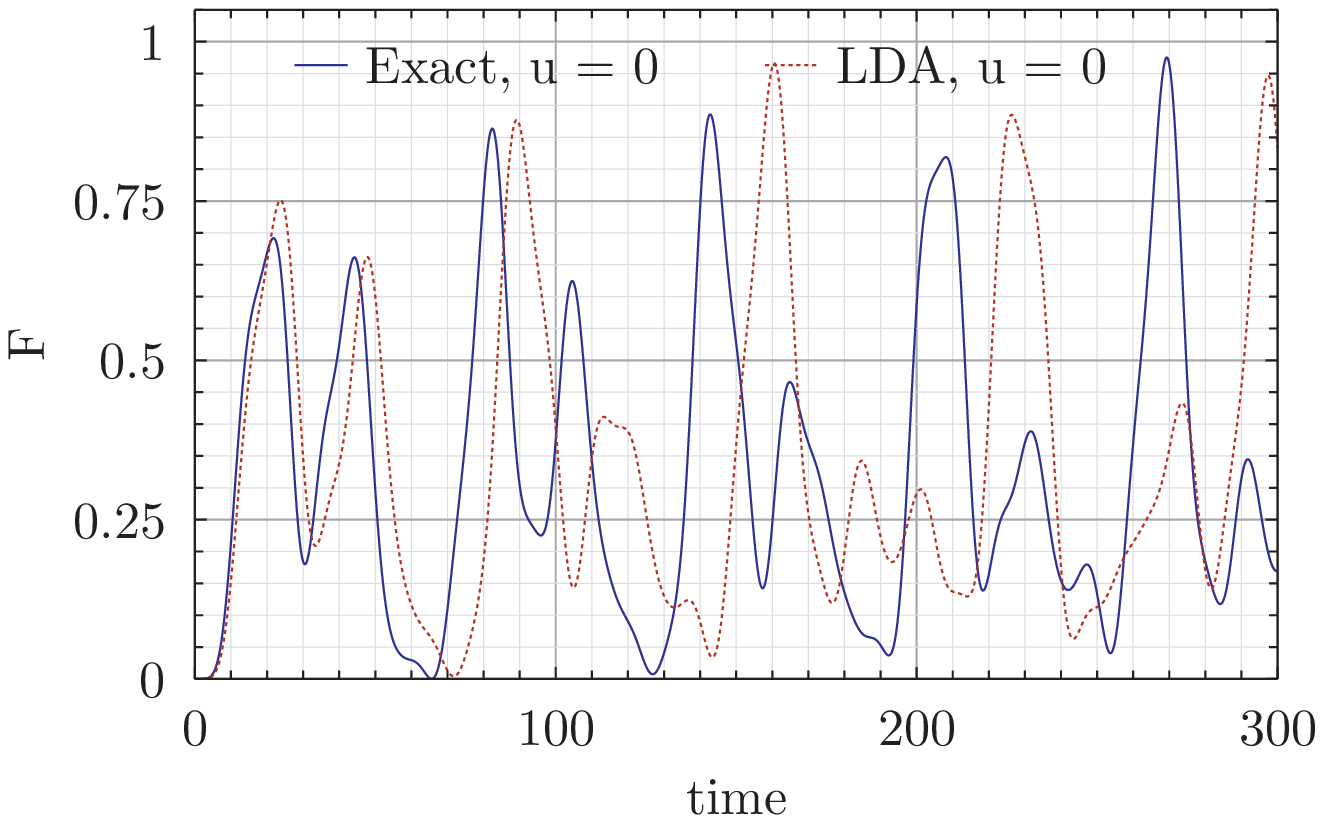}
\includegraphics[scale=\figscale]{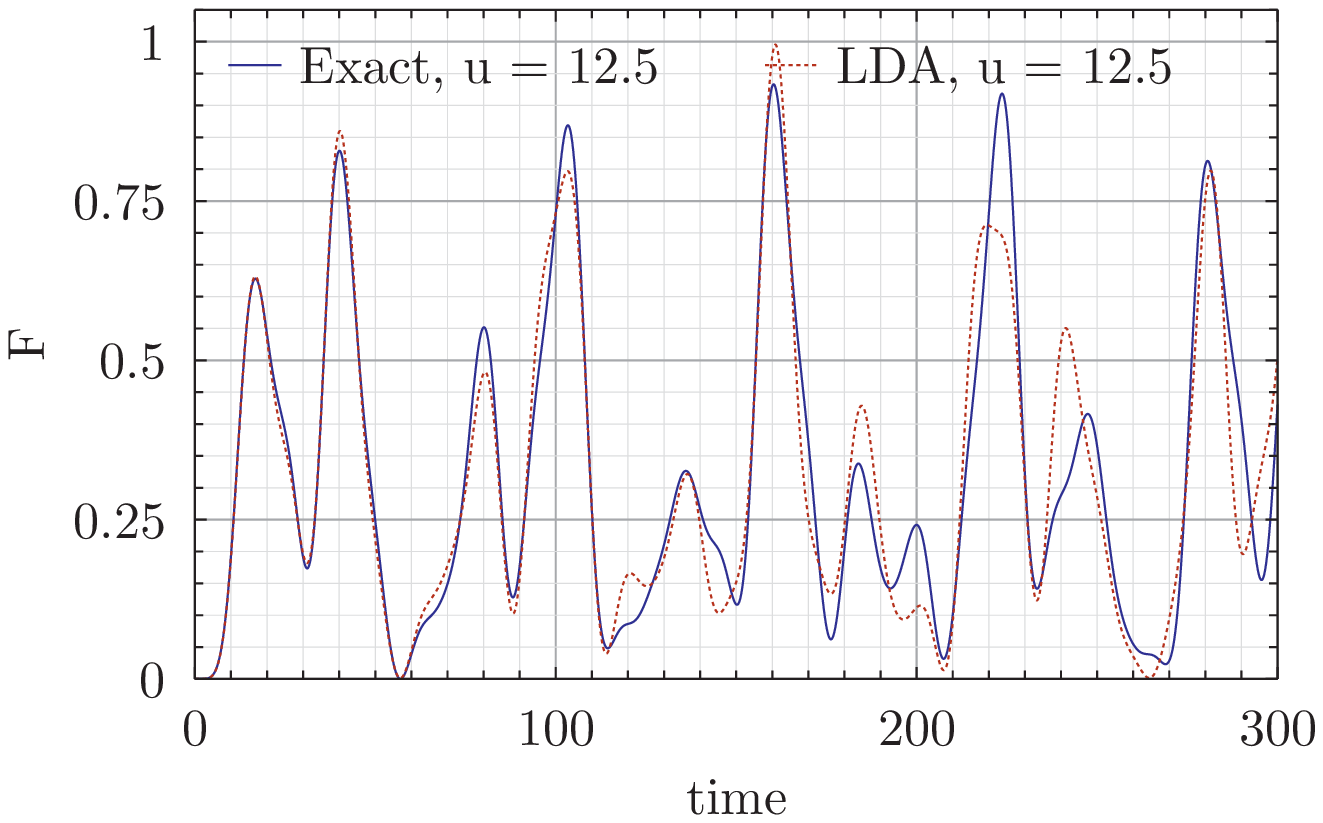}
\includegraphics[scale=\figscale]{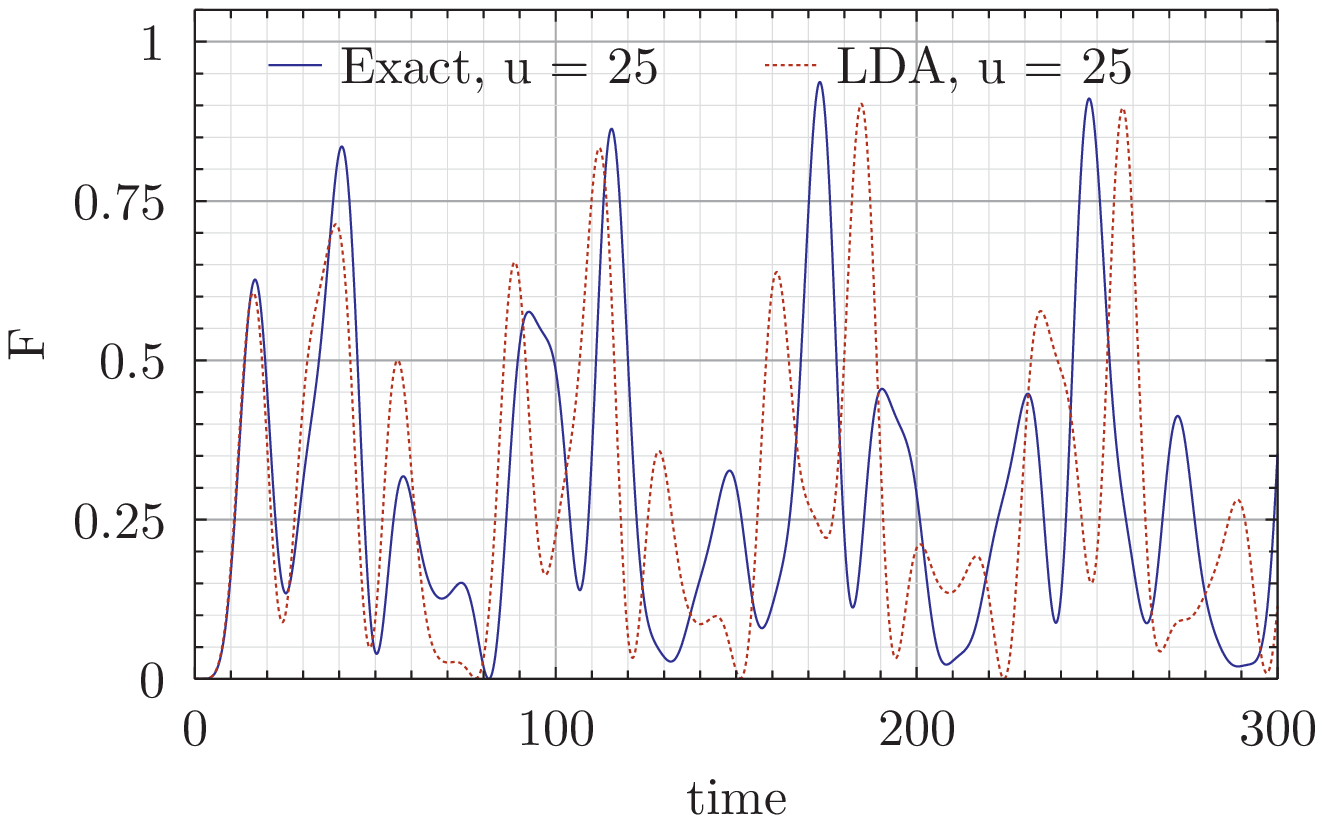}
\includegraphics[scale=\figscale]{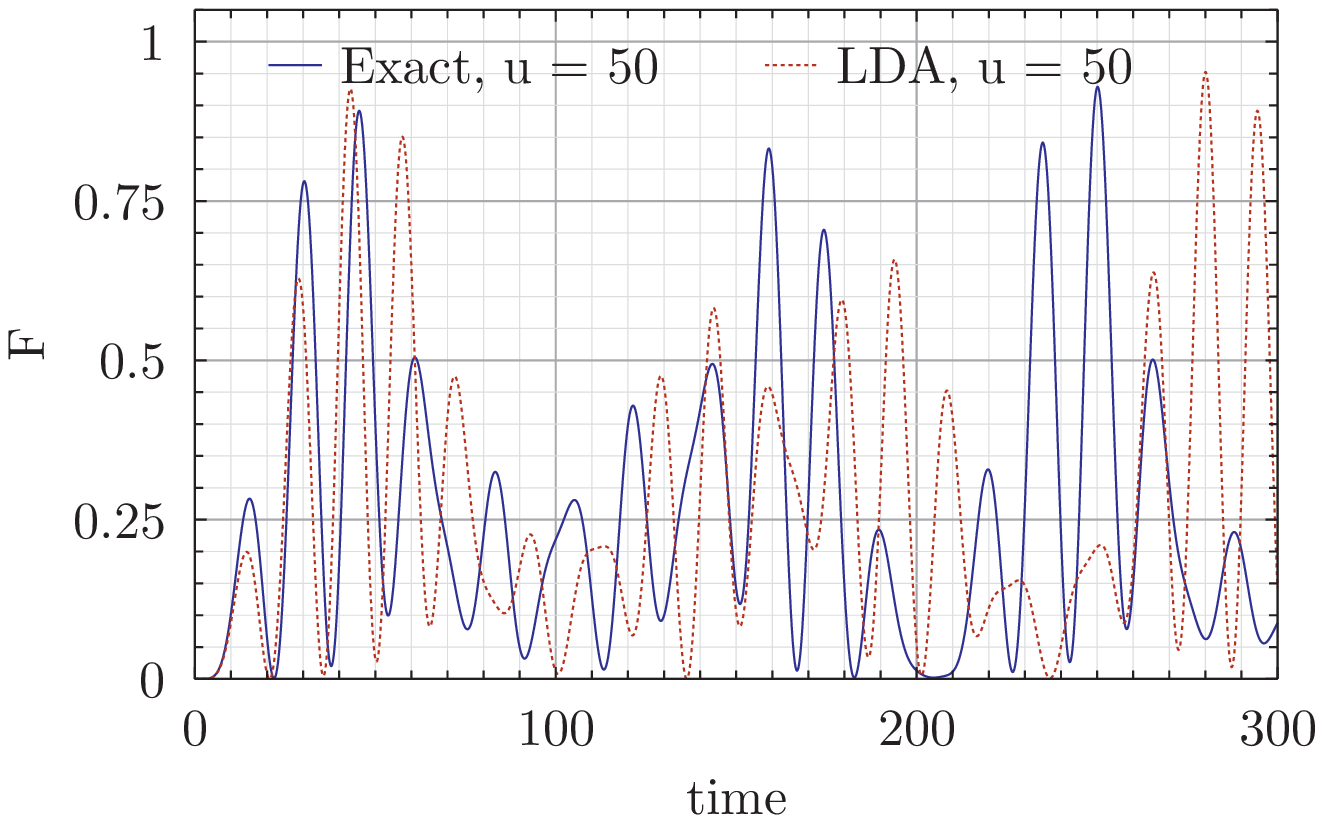}
\end{minipage}
\caption{Quantum state transfer fidelity, $F$, for an $N=4$ two-component Fermi system with $J_i$ obtained from the 
potential in Eq.~\eqref{AP} for $u=0$ (first row), $u=12.5$ (second row), $u=25$ (third row), and $u=50$ (fourth row) 
where $u$ is in units of $\epsilon$. The other potential parameter is $V_0=50$ for all panels.
The (blue) solid line corresponds to the $J_i$ coefficients obtained from the exact calculation and
the (brown) dashed line corresponds to the LDA approach. Time is measured in units 
of $\hbar/\epsilon$.}
\label{fig:APFidelityN=4}
\end{figure}

\section{Quantum transport properties}\label{qst}
In order to understand some potential effects that could be 
implied by using the LDA instead of the exact solutions for the 
exchange couplings, we now consider a dynamical protocol 
that has recently been discussed in the context of strongly 
interacting systems in 1D. As mentioned earlier, one 
may indeed map these systems into effective spin models
\cite{volosniev2013,deuret2014}. This can be done for both
Fermi systems \cite{volosniev2013,deuret2014,levinsen2014}
and Bose systems \cite{volosniev2014,pietro2015}. To keep 
the discussion concise, we will mainly consider the example of the 
two-component Fermi system here. In that case, the spin 
mapping is into the famous Heisenberg spin-$\tfrac{1}{2}$ 
model whose 
Hamiltonian (up to a constant energy shift) is given by
\begin{equation}
 H_{s} =  \sum_{j=1}^{N-1}  J_j \bm S^j \cdot \bm S^{j+1},
 \label{eq:spinHam}
\end{equation}
where $\bm S^j = \frac{1}{2} \bm \sigma^j$ is a spin operator where $\bm \sigma^j = (\sigma_x^j, \sigma_y^j, \sigma_z^j)$
is the vector of the Pauli matrices. Here the nearest-neighbor 
interaction coefficients are related very simply to the 
local exchange coefficients and the coupling constant 
as $J_{i} \equiv - \frac{\alpha_i}{g}$. This is another
justification for using the term 'local exchange 
coupling' for the $\alpha_i$ above, i.e. that under 
the spin mapping they appear as nearest-neighbor 
couplings in what is equivalent to a spin chain
Hamiltonian. For Bose systems, one may have more 
involved spin models as the coefficients for 
$x$-, $y$-, and $z$- direction spin operators
are not necessarily the same (see Ref.~\cite{volosniev2014} 
for further details). Below, we will make one 
detour from the uniform Heisenberg spin model in 
Eq.~\eqref{eq:spinHam} in order to consider the 
so-called XX model which is a special case of 
the Hamiltonian in Eq.~\eqref{eq:spinHam} 
where all the terms with $z$-component operators 
are eliminated.

\begin{figure}
\begin{minipage}{0.9\linewidth}
\includegraphics[scale=\figscale]{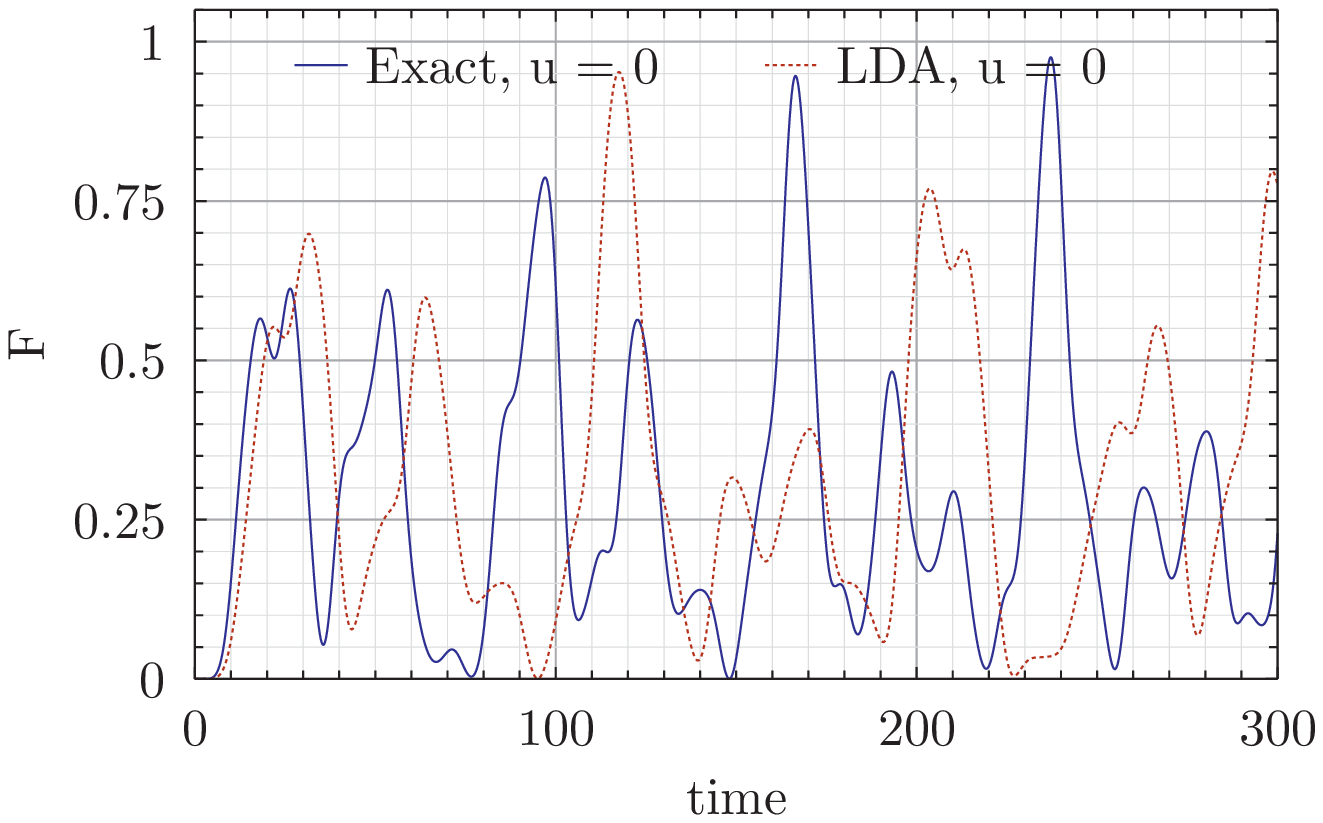}
\includegraphics[scale=\figscale]{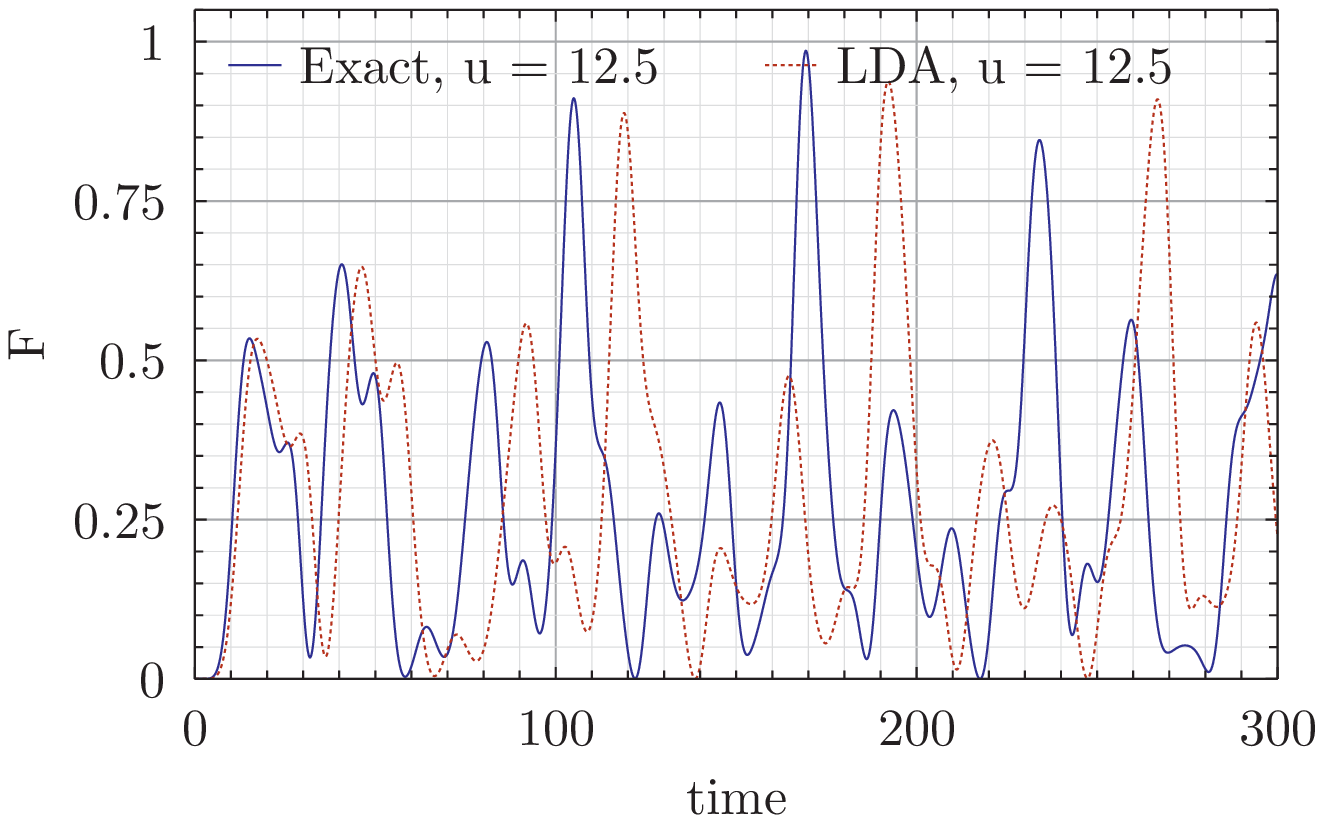}
\includegraphics[scale=\figscale]{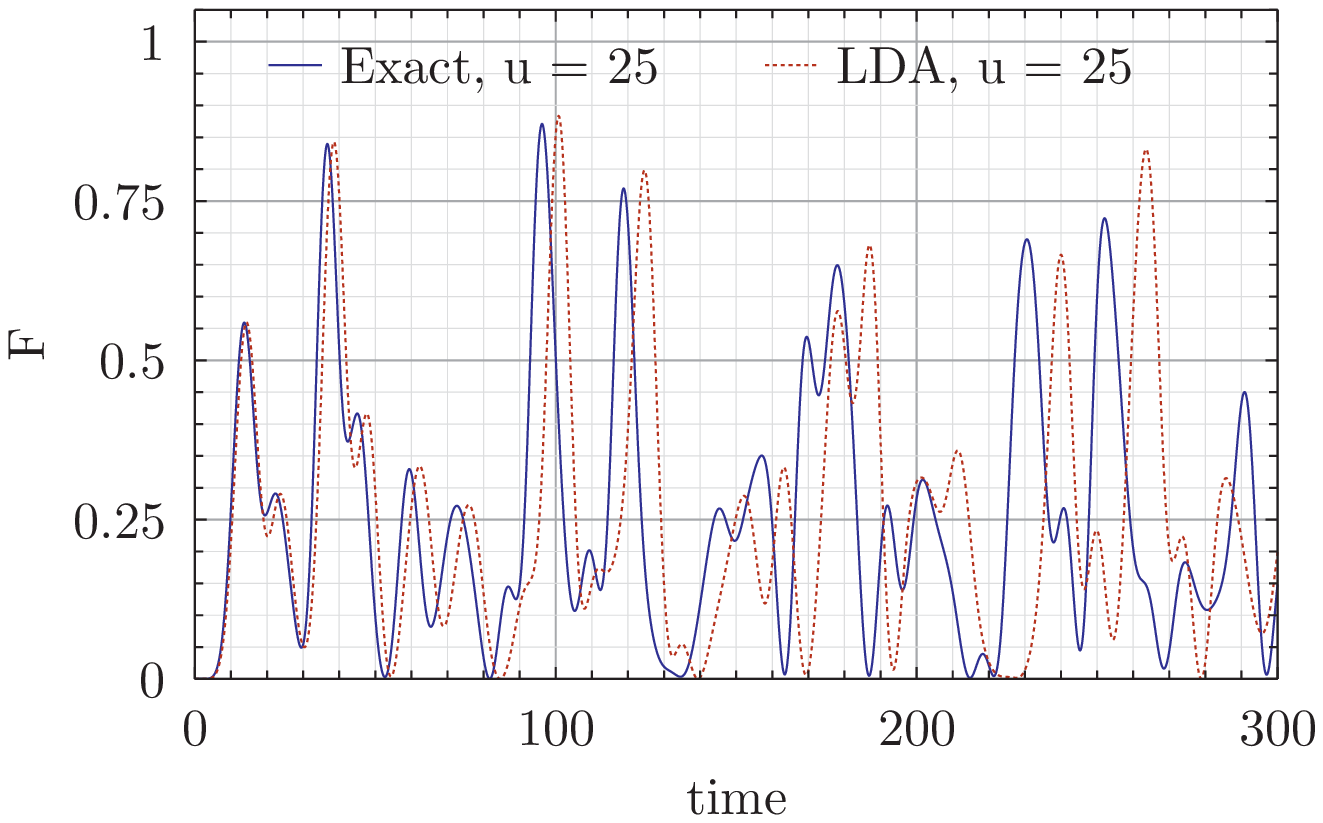}
\includegraphics[scale=\figscale]{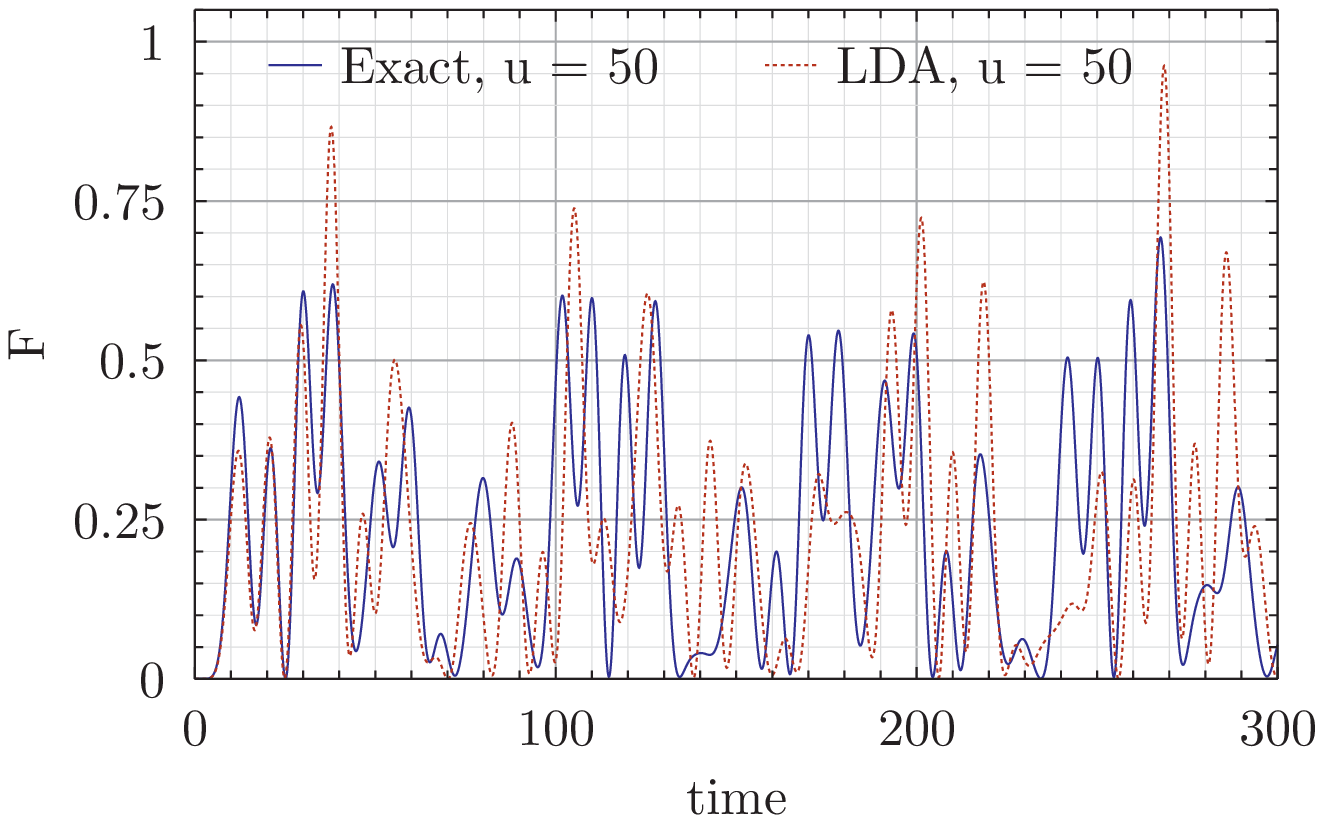}
\end{minipage}\hfil
\caption{Same as Fig.~\ref{fig:APFidelityN=4} but for $N=5$.}
\label{fig:APFidelityN=5}
\end{figure}

Spin-$\tfrac{1}{2}$ chains have been proposed as 
media for quantum state transfer about a decade ago 
\cite{bose2003, bose2008, christandl2005}. The quantum 
state transfer in these chains essentially corresponds 
to flipping a single spin at one end of the chain 
and then dynamically evolving the state such that one 
may project onto the state at which the single spin 
flip has reached the other end for all subsequent times.
The probability of finding the flipped spin at the other
end is known as the fidelity of the quantum state transfer
and if it reaches unity at some subsequent time we say
that the system supports perfect quantum state transfer
\cite{bose2008}. In the original proposal \cite{bose2003}
a spin chain with constant $J_i=J$ was considered and 
it was shown that the fidelity could not reach unity 
in this case. Later studies demonstrated that if the 
coefficients are chosen as $J_i=J_0\sqrt{N(N-i)}$ with 
$J_0$ some overall constant, then perfect state transfer
is possible (although only within the so-called XX model). 
One would then have an ideal communications channel for quantum 
information. However, it turns out to be exceedingly difficult
to produce a system that fulfills these requirements on 
$J_i$. In Ref.~\cite{volosniev2014} Volosniev {\it et al.} 
proposed strongly interacting 1D atomic systems as a possible 
realization of perfect state transfer and found that the 
potential in Eq.~\eqref{AP} can give rise to perfect state
transfer in the XX model for $u=12.5$. We will return to this below.
First we need to define the spin state space and the fidelity 
that we consider. 
For a spin chain with one impurity (a single spin that is flipped) 
and $N-1$ majorities (in other words $N_\uparrow = 1$ and $N_\downarrow = N - 1$) 
it is natural to use the basis of spin functions $\left \{ \mket{\uparrow\downarrow\downarrow\ldots\downarrow}, \mket{\downarrow\uparrow\downarrow\ldots\downarrow}
, \mket{\downarrow\ldots\downarrow\uparrow\downarrow}, \mket{\downarrow\ldots\downarrow\downarrow\uparrow}\right \}$. In this
basis we define the fidelity of the quantum state transfer which is a
function of time and is given by
\begin{equation}
 F(t) = \left | \mbra{\downarrow\ldots\downarrow\downarrow\uparrow} e^{-iH_{s}t/\hbar} \mket{\uparrow\downarrow\downarrow\ldots\downarrow} \right |^2.
 \label{eq:fidelity}
\end{equation}
The fidelity can be straightforwardly interpreted as the Hamiltonian 
acting on the initial state on the right (with the single flipped 
spin on the left edge of the systems) and then projection on the 
final state on the left (with the single flipped spin on the 
right edge). In practice, one expands the initial state on the 
set of eigenstates of the Hamiltonian, then constructs the 
state at all later times, and then projects onto the final 
state which also has some expansion in terms of the 
eigenstates of the Hamiltonian. Notice that the overall time
scale of the transfer depends on $\alpha_i$ and on $g$. In 
the strict limit where $1/g=0$, there is no dynamics as 
the particles are completely impenetrable and all orderings
are eigenstates (all of which are degenerate). However, in 
the more realistic case where $g$ is large but finite, our
effective spin models work to linear order in $1/g$ and can 
thus be used to study dynamics in the strongly interacting regime.
The timescale of transfer will then depend linearly on $g$. 
One may think of the state transfer process as a set of 
subsequent flips of pairs of spins along the chain. Each of 
these local exchanges depend on $\alpha_i$, i.e. if $\alpha_i$
is large this happens fast and vice versa. We thus see that 
large barriers will suppress the transfer as expected. 
However, the process can depend rather delicately on the 
actual values of the local exchanges as we will now demonstrate.
For instance, in the exact results we may see slow suppression 
of the local exchange with barrier height as in Fig.~\ref{fig:DWcomp}
in regimes where LDA would give exactly zero (when the chemical 
potential is below the barrier). This is one source of 
error in the LDA that could carry into a dynamical protocol 
like state transfer in a severe way.

\begin{figure}
\begin{minipage}{0.9\linewidth}
\includegraphics[scale=\figscale]{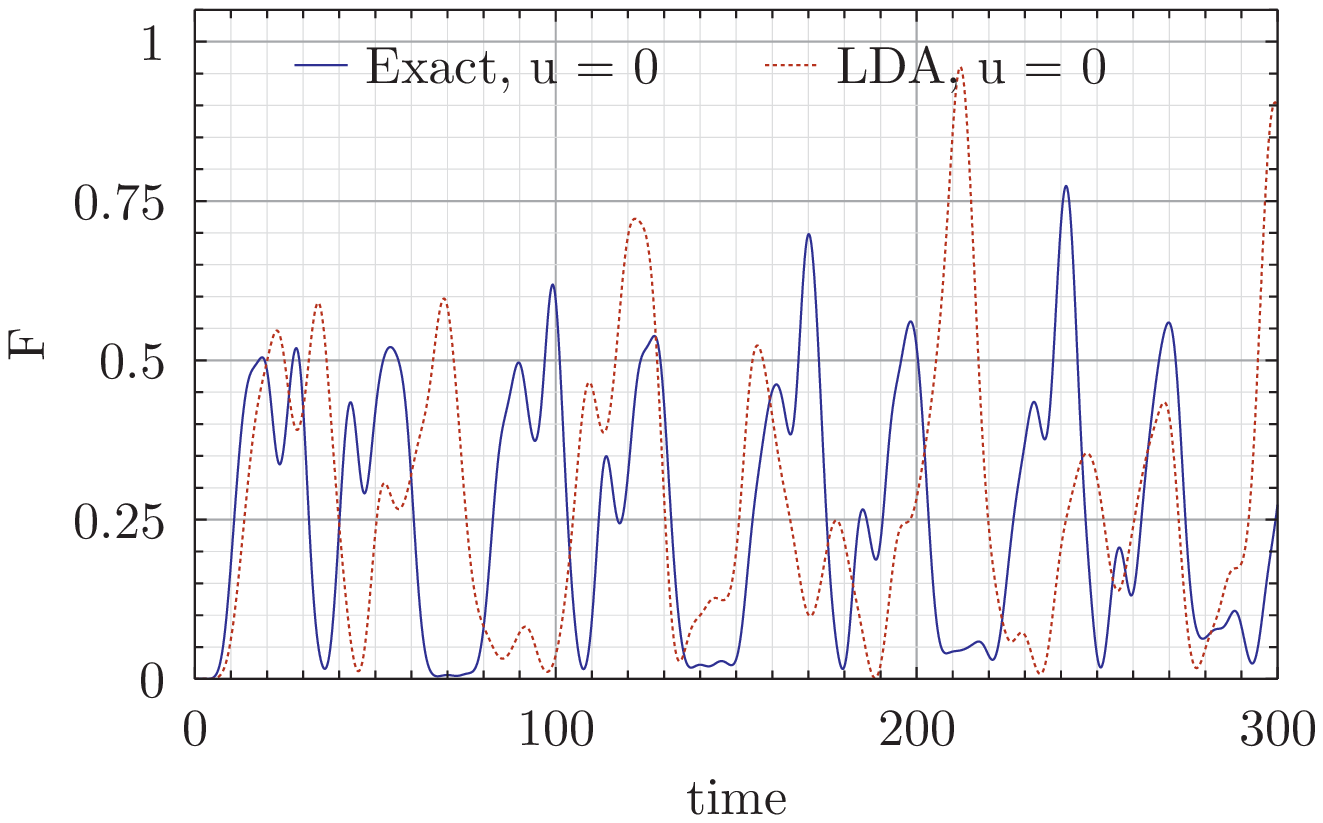}
\includegraphics[scale=\figscale]{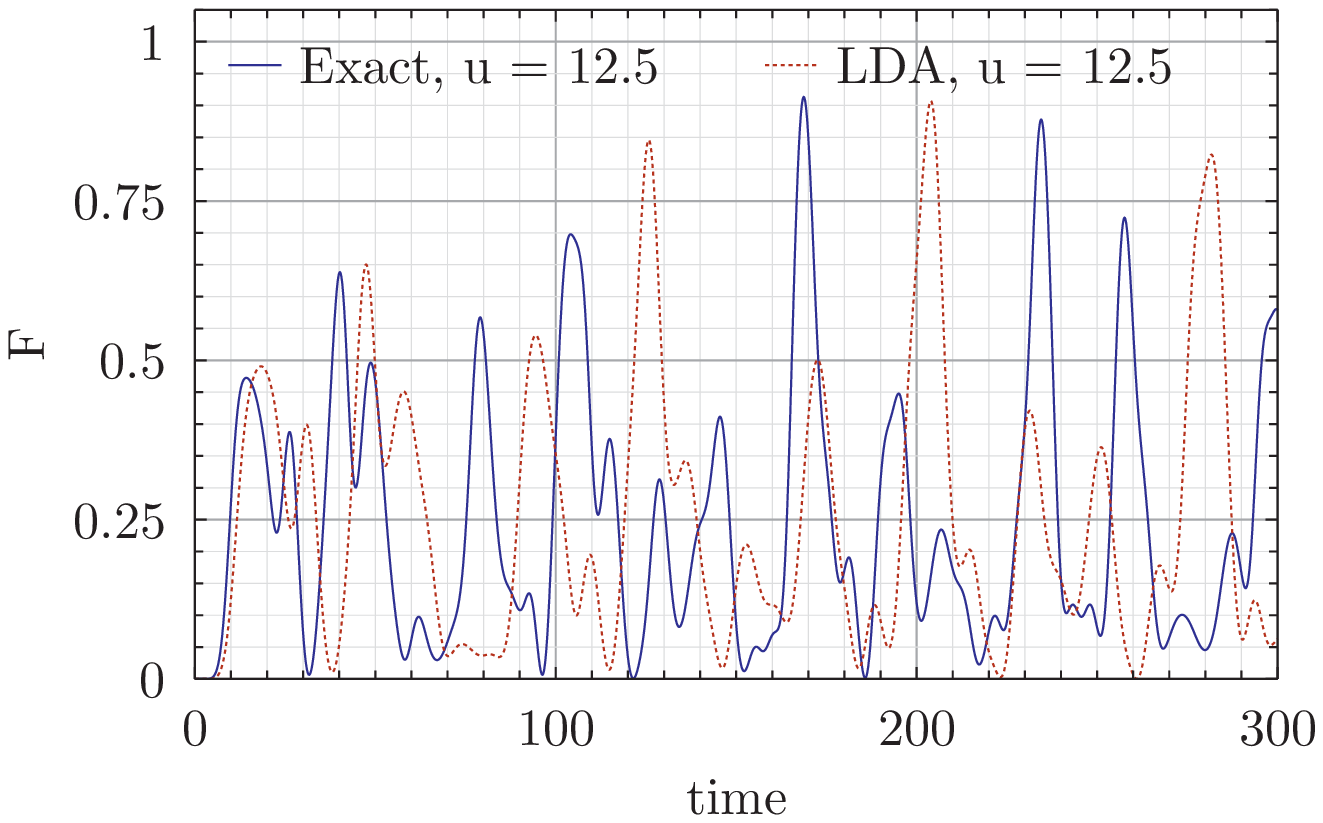}
\includegraphics[scale=\figscale]{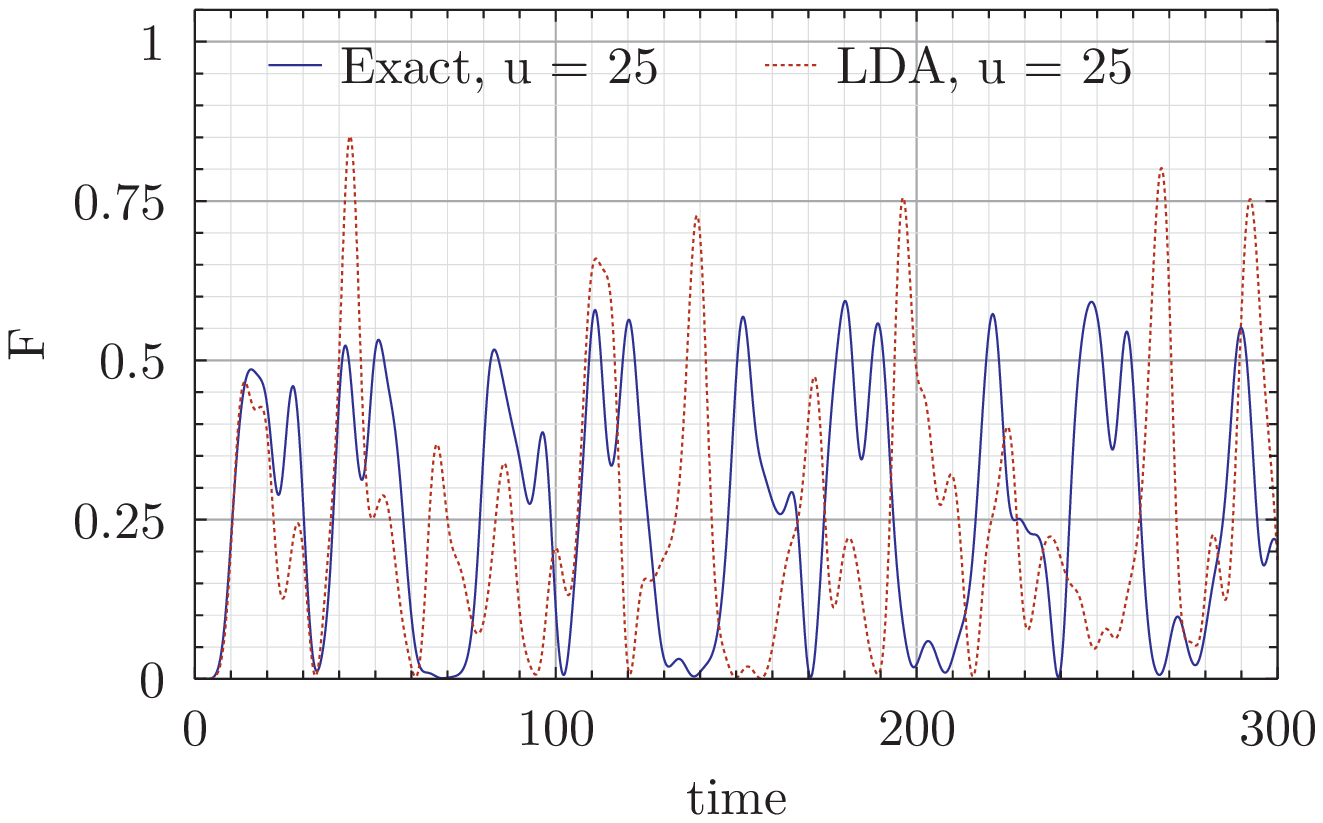}
\includegraphics[scale=\figscale]{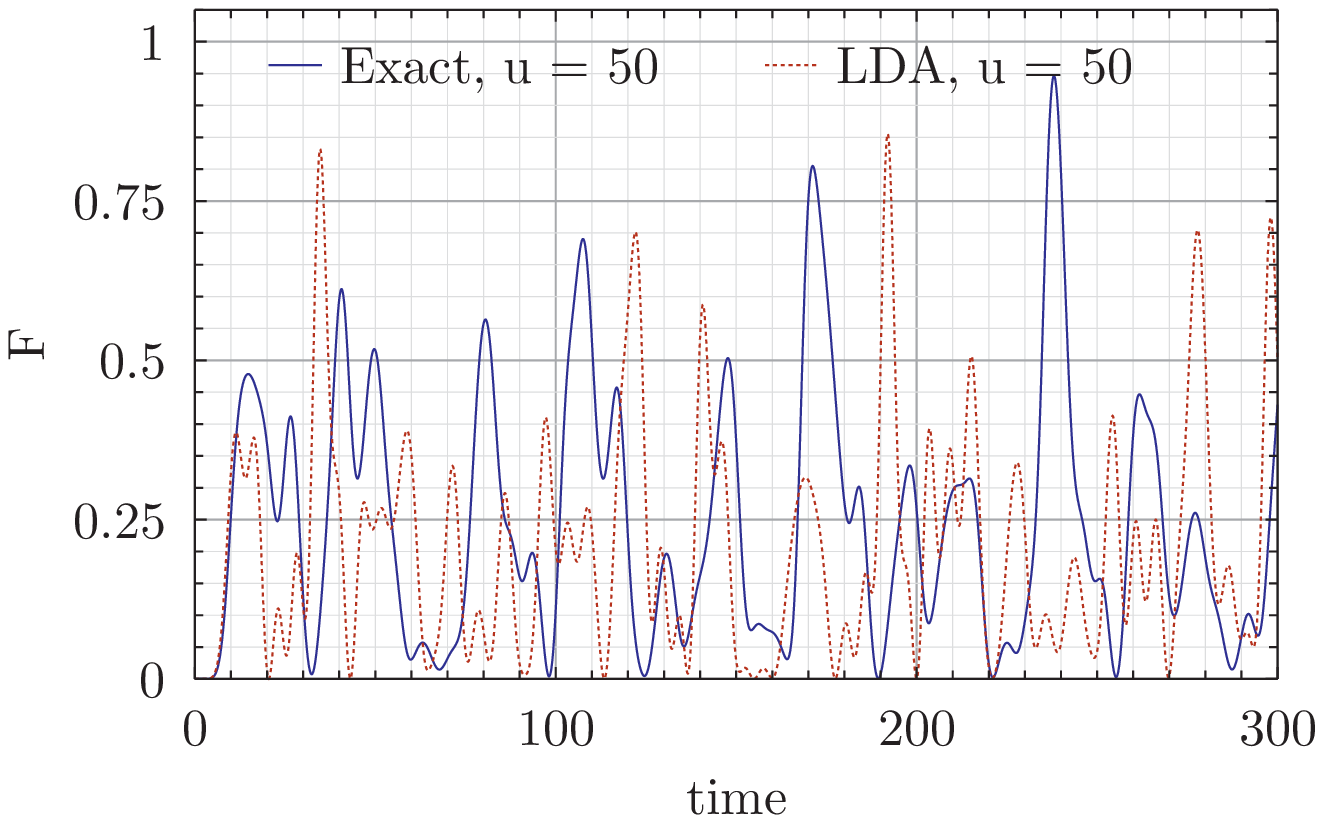}
\end{minipage}\hfil
\caption{Same as Fig.~\ref{fig:APFidelityN=4} but for $N=6$.}
\label{fig:APFidelityN=6}
\end{figure}

Here we plot the fidelity of Eq.~\eqref{eq:fidelity} as a function of
time for different potentials of the form given in Eq.~\eqref{AP} for 
the cases with $N=4$ (Fig.~\ref{fig:APFidelityN=4}), 5 
(Fig.~\ref{fig:APFidelityN=5}), and 6 (Fig.~\ref{fig:APFidelityN=6}) particles, 
respectively. All the three cases we present show the characteristic 
oscillatory behavior of state transfer fidelities as function of 
time \cite{volosniev2014}. However, one does indeed notice that 
the oscillations become more prominent and more irregular as 
$N$ increases (system size or chain length in the 
spin chain language), and also as the control parameter for
the central barrier $u$ is increased. 
Perfect state transfer
is never achieved for these parameters in a Heisenberg model of the type
in Eq.~\eqref{eq:spinHam} 
without an external magnetic field. This is in accordance
with previous results~\cite{christandl2005,christandl2004,nikolopoulos2004}. 
The calculations demonstrate that even when there are only small 
differences in the exact and LDA results, the fidelity can show
large variations particularly for longer time intervals. This is 
probably due to an accumulation of phase factors in the system over time that tends
to drive the exact and LDA results apart. However, we do see 
some instances of very good agreement as for instance in the 
second row of Fig.~\ref{fig:APFidelityN=4} where $u=12.5$ 
and $V_0=50$ in Eq.~\eqref{AP}. 
Even in this case, noticeable 
differences between exact and LDA results are seen, 
but the overall agreement is very good. 
For the larger 
particle numbers in Figs.~\ref{fig:APFidelityN=5} and 
\ref{fig:APFidelityN=6} we may even notice a clear 
tendency for the LDA results to predict large fidelity
spikes that are either at different times as compared to 
similar spikes using the exact result, or in some 
instances the LDA shows spikes where the exact results 
have none (see for instance the fourth row in
Fig.~\ref{fig:APFidelityN=5} or the third row in Fig.~\ref{fig:APFidelityN=6}).
We thus conclude that in most cases, the LDA results can 
provide large deviations from the exact results in a 
dynamical process such as quantum state transfer.

\begin{figure}
\centering
\includegraphics[scale=0.65]{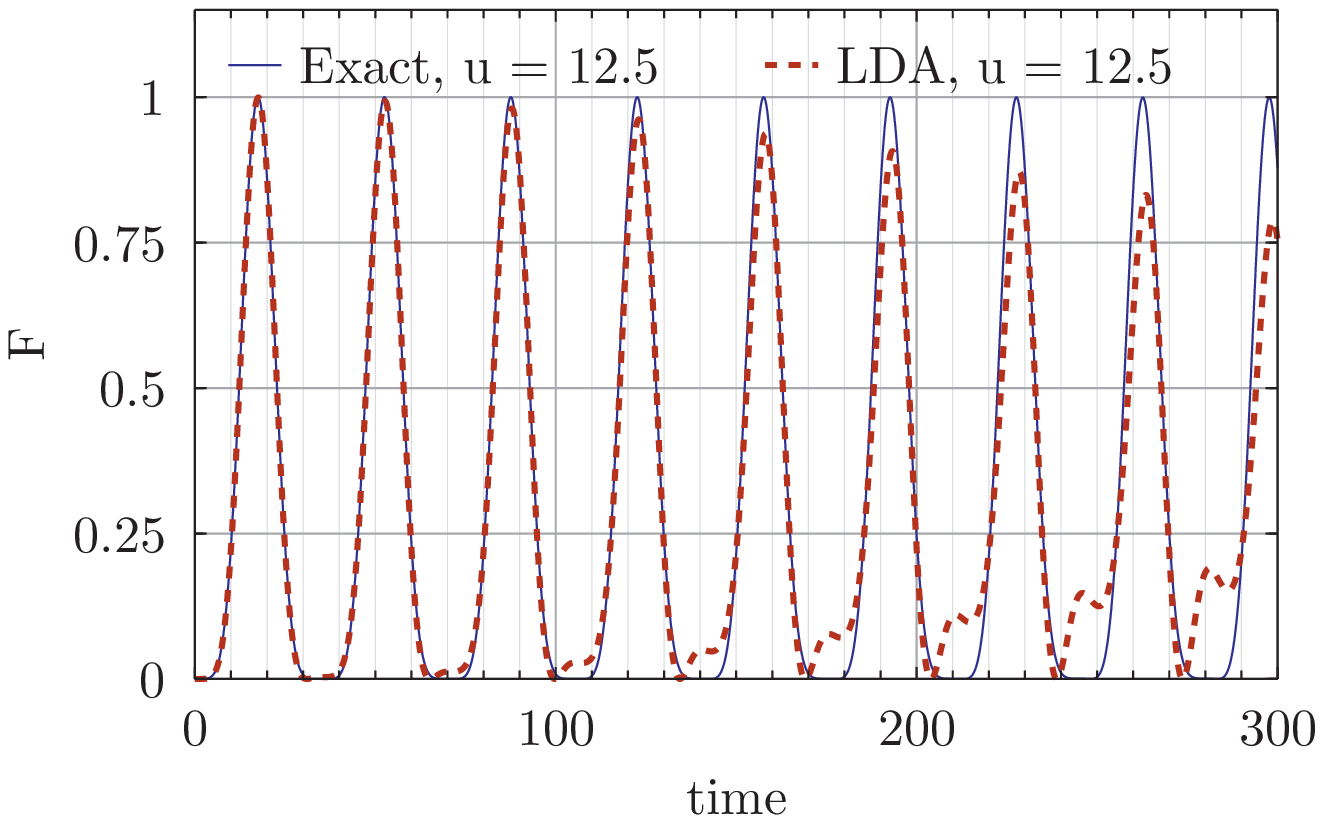}
\caption{Quantum state transfer fidelity, $F$, for an $N=4$ two-component Fermi 
system with $J_i$ obtained from the 
potential in Eq.~\eqref{AP} for $u=12.5$ and $V_0=50$ (in units of $\epsilon$) 
for the case where the $z$-components in Eq.~\eqref{eq:spinHam}
can be discarded so that we are in the so-called Heisenberg XX spin model. In this case perfect state transfer
can be achieved \cite{volosniev2014}.
The (blue) solid line corresponds to the $J_i$ coefficients obtained from the exact calculation and
the (brown) dashed line corresponds to the LDA approach. Time is measured in units 
of $\hbar/\epsilon$.}
\label{fig:APPerfTransferN=4}
\end{figure}

In closing this section, we want to consider how well 
the LDA results do in the case where perfect state 
transfer is achieved. This can be achieved with parameters
corresponding to the second row in Fig.~\ref{fig:APFidelityN=4}
in a setup where we can discard the $\bm S_z^j \bm S_z^{j+1}$ 
interaction in the Hamiltonian in Eq.~\eqref{eq:spinHam} 
and thus reduce the problem to the Heisenberg XX model.
This can be realized by either using a tailored external magnetic
field applied to the system or in specific models with 
strongly interacting two-component Bose systems \cite{volosniev2014}.
In Fig.~\ref{fig:APPerfTransferN=4} we compare the fidelities computed
using the
exact and the LDA values of the exchange coupling coefficients. We see that 
the perfect transfer is indeed obtained for the values 
calculated via Eq.~\eqref{eq:alphaExact}. However, the fidelity of 
the spin transfer decreases with time for the coefficients 
obtained via Eq.~\eqref{eq:alphaLDA}.
The relative differences in the $\alpha_i$ exchange coefficients 
are no more than $2$ percent when comparing the exact to the LDA
method. However, we still see that 
it affects the quantum transfer properties on 
longer time scales significantly.

\section{Summary and discussion}\label{summary}
We have considered strongly interacting two-component systems in one dimension 
held in place by an external confinement. Such systems have an effective Hamiltonian
that can be completely specified by computing a set of local exchange coefficients
which may be interpreted as nearest-neighbor spin exchange interactions when the 
system is mapped onto a spin model Hamiltonian of the Heisenberg type. Computing 
these exchange coefficients as accurately as possible is an important yet also 
computationally difficult task. In the present paper we have explored two different
approaches. One is a 'brute force' calculation of a multi-dimensional integral which 
gives the exact result (but is prohibitive for larger particle numbers) and the 
other is an approach inspired by the local density approximation (LDA) that could in 
principle be used to reduce the computational complexity. While previous studies 
have shown that the local density approach can be accurate at the level of a 
few percent for the case of a harmonic oscillator potential, we explore more 
complicated geometries consisting of two instances of a double-well potential. 

Our findings demonstrate that while the LDA does rather
well with potentials that resemble a harmonic oscillator, the 
exchange couplings do not have the right qualitative and quantitative behavior 
in the LDA when there are significant barriers as is typical of a double-well
potential. In particular, the LDA cannot capture the right quantum tunneling 
processes across such barriers and may thus leave out important effects.
We have shown that for small systems this can lead to an underestimation of 
exchange by the LDA for potentials which have a more interesting structure than the simple 
single-well harmonic oscillator. 
We expect this observation to have influence on experiments
with small system sizes in 1D optical lattices. There one typically also has an external 
overall smooth potential (approximately harmonic in shape). Thus, the potential seen by 
the particles is the superposition of harmonic trap and optical lattice, i.e. a 'smiling'
lattice potential. 
This is a very structured potential and in the strongly interacting 
regime one could be in dire straits with a simple LDA approach as compared to the exact exchange
couplings. In order to test out the LDA in a concrete physical process, we considered
quantum state transfer of single spin flips in systems with four, five, and six
particles. Here we found that while the LDA performs reasonably for some double-well
realizations there is amplification of the deviations of the LDA compared
to the exact results in the transfer fidelity that can be significant and lead to large 
errors in both maximum fidelity values and the specific times at which these are
attained. 

The current study has concentrated on small particle numbers where very accurate
results can be obtained for the local exchange coefficients using multi-dimensional 
integration so that a comparison between exact and LDA results is possible for 
arbitrary potentials. Incidentally, as we discussed in the introduction, the 
system sizes used here are also of great current experimental interest in cold 
atoms. One would, however, like to study how these results scale to larger particle
numbers. This most likely requires alternative approaches not only to the exact
formula but also to the LDA formula. In the current implementation it depends
on the density of the system which is not easy to compute for arbitrary potentials. 
One may thus pursue an agenda of finding an alternative LDA method that obtains 
the density by some other and computationally much simpler approach, and simultaneously
explore how to get a computational reduction of the exact formula. 

The authors gratefully acknowledge discussions with and feedback from A.~G. Volosniev, 
M. Valiente, D.~Petrosyan, N.~J.~S. Loft, A.~E. Thomsen, and L.~B. Kristensen. This 
work was supported by the Danish Council for Independent Research DFF Natural Sciences 
and the DFF Sapere Aude program.

\end{document}